\documentclass[reprint,superscriptaddress,amsmath,amssymb,amsfonts,aps,prb]{revtex4-1}

\usepackage{graphicx}
\usepackage{dcolumn}
\usepackage{color}
\usepackage{bm}
\usepackage{hyperref}
\usepackage[table]{xcolor}

\begin{document}

\title{Random iron-nickel alloys: From first principles to dynamic spin-fluctuation theory}

\author{G.~V. Paradezhenko}
\affiliation{Skolkovo Institute of Science and Technology, Moscow 121205, Russia}

\author{D. Yudin}
\affiliation{Skolkovo Institute of Science and Technology, Moscow 121205, Russia}

\author{A.~A. Pervishko}
\email{a.pervishko@skoltech.ru}
\affiliation{Skolkovo Institute of Science and Technology, Moscow 121205, Russia}

\date{\today}

\begin{abstract}
We provide a systematic analysis of finite-temperature magnetic properties of random alloys Fe$_x$Ni$_{1-x}$ with the face-centered-cubic structure over a broad concentration range $x$. By means of the spin-polarized relativistic Korringa-Kohn-Rostoker method we calculate the electronic structure of disordered iron-nickel alloys and discuss how a composition change affects magnetic moments of Fe and Ni and the density of states. We investigate how the Curie temperature depends on Fe concentration using conventional approaches, such as mean-field approximation or Monte Carlo simulations, and dynamic spin-fluctuation theory. Being devised to account for spin fluctuations explicitly, the latter method shows the best fit to experimental results.
\end{abstract}

\maketitle

\section{Introduction}

Transition metal binaries have seen a surge of interest for many decades due to their unique magnetic characteristics, and disordered iron-nickel alloys are a typical example. In practice, varying elemental composition serves as a means to tune their magnetic properties and crystalline structure \cite{Swartzendruber1991,Mao2006,Cacciamani2010, Glaubitz2011,Xiong2011}. Among the most remarkable effects which iron-nickel alloys are known to exhibit is the Invar effect that is manifested in the lack of thermal expansion over a wide temperature range \cite{Kakehashi1981,Moruzzi1990, Wasserman1990, Abrikosov1995,Schroter1995,Wassermann1991,Van1999, Khmelevskyi2004}. Also noteworthy is a sharp drop of saturation magnetization for Invar composition range \cite{Asano1969,Window1973,Sedov1998,Glaubitz2011}. Equally noticeable is the studying of permalloy demonstrating very high magnetic permeability and relatively small magnetocrystalline anisotropy and magnetostriction \cite{Bozorth1953,Ferguson1958,Pfeifer1980,Hatafuku1983,James2000,Yin2006,Ustinovshikov2013,Waeckerle2020}. All these factors stimulated the efforts in detailing the physical mechanisms underlying Fe-Ni alloys.

From a theoretical viewpoint, a standard approach for studying disordered systems relies on the Korringa-Kohn-Rostoker method within the coherent potential approximation (KKR-CPA) based on density functional theory in its local-spin density approximation (LSDA) \cite{Faulkner1982,Feder1983,Strange1989,Akai1989,Dederichs1991,Ebert1992}. The CPA was subsequently generalized to account for short-range effects---the nonlocal coherent-potential approximation (NL-CPA)~\cite{Rowlands2003}. For random alloys without short-range order, the latter yields the results close to the ones obtained within standard CPA~\cite{Rowlands2005} (see a detailed review on NL-CPA in Ref.~\cite{Row09}). To address a delicate interplay between many-body effects and disorder, various schemes have been further combined, including LSDA$+U$ \cite{Ebert2003} and LSDA with dynamical mean-field theory (DMFT) \cite{Minar2005,Sipr2008, Braun2010,Minar2014,Poteryaev2016}. It should be noted that it is  also possible to examine the random substitutional alloy within a fully polymorphous supercell description~\cite{FUM01, Popescu2010}.

Despite the significant progress in developing various numerical tools that allow one to describe magnetic properties of alloys at finite temperatures, a reliable estimate of the Curie temperature in these compounds still remains a challenging problem. Most commonly, one adopts various modifications of mean-field approximation (MFA) and Monte Carlo simulations, where the structure features are introduced via magnetic exchange parameters of a classical Heisenberg model, as evaluated from first principles neglecting the fluctuation contributions \cite{Mano1977,Liechtenstein1987, Halilov1998,Turek2006,Takahashi2007,Evans2020}. However, a proper treatment of thermodynamic properties of a given compound should inevitably take into account many-body effects originating from both local and long-wave spin fluctuations. Note that the effect of finite temperature on magnetic fluctuations was discussed in the framework of the disordered local moment (DLM) method~\cite{Pindor1983,Gyorffy1985,Staunton1992}, giving relatively good agreement with experimentally accessible quantities~\cite{Staunton2004,Kudrnovsky2012}. In particular, the importance of longitudinal spin fluctuations in describing high temperature properties of selected iron- and nickel-based systems within developed spin-fluctuation models had been recently emphasized in Refs.~\cite{Hasegawa1981,Uhl1996,Rosengaard1997,Kakehashi2002,Shallcross2005,Ruban2007,MD12,Grebennikov2015,Ruban2016,Ruban2017,Pan2017,MR18}. This led thereafter to the idea of accounting for combined contribution from the spin fluctuations and thermal lattice vibrations on the basis of the alloy analogy model~\cite{Ebert2015}.

In this paper, we aim to complement previous findings by investigating magnetic properties of Fe$_x$Ni$_{1-x}$ random alloys in dependence on specific iron concentration $x$, and by taking account of on-site and nonlocal interactions. On first-principles ground, we report on the concentration dependence of disordered Fe$_x$Ni$_{1-x}$ in face-centered structure, limiting our consideration to $0.1\leq x\leq 0.6$. In iron-nickel alloys containing about 40\% of Ni, instability of the magnetic moment with respect to a volume change develops towards a martensitic transition that brings the system to the body-centered-cubic structure \cite{Asano1969,Glaubitz2011}. We evaluate the element-resolved and average spin magnetic moments of Fe$_x$Ni$_{1-x}$ binary alloys along with material spin-resolved densities of states. These results are further used to benchmark a variety of numerical methods, including MFA, Monte Carlo simulations, and dynamic spin-fluctuation theory (DSFT), against experimental measurements of the Curie temperature. As opposed to other methods, DSFT correctly reproduces the dependence
of the Curie temperature upon increasing $x$.

\section{Computational details}

The numerical calculations are carried out for random alloys of Fe$_x$Ni$_{1-x}$ with face-centered-cubic structure corresponding to the $Fm \bar{3} m$ space group with the lattice parameter $a$ increasing as a function of the alloy composition $x$~\cite{Swartzendruber1991,Glaubitz2011}. The experimental alloy lattice constants utilized throughout the paper are listed in Table~\ref{tab:latmom}.

We address the electronic structure of random binaries within the spin-polarized relativistic Korringa-Kohn-Rostoker Green's function formalism as implemented in the Munich SPR-KKR package~\cite{SPRKKR, SPRKKR1}, where the effect of substitutional disorder is taken into account by using CPA. The presented results are obtained in the spin-polarized scalar-relativistic mode, where the atom magnetization is oriented along the $z$-crystallographic axis. To achieve self-convergence, we apply the BROYDEN2 algorithm~\cite{Bro65,DS96} and Vosko-Wilk-Nusair parametrization for the exchange-correlation potential~\cite{VWN}. For transition metals, the angular momentum of the Green's function cutoff $l_{max}=2$ is considered generally sufficient; however, to validate the electronic structure that is subsequently utilized as an input in DSFT calculations we assess the influence of increased a cutoff value on electronic properties of the alloy by employing cutoff up to $l_{max}=3$ for each atom~\cite{Stefanou1987,Mankovsky2013,Ebert2015}. During the self-consistent potential study, we utilize $22 \times 22\times 22$ $k$-point mesh, while the subsequent calculations of the density of states (DOS), magnetic moments and Heisenberg exchange coupling strength are performed with $57 \times 57 \times 57$ mesh. The Liechtenstein-Katsnelson-Antropov-Gubanov formalism~\cite{Liechtenstein1987} is adapted for estimating the exchange coupling parameter by employing a cluster with tripled lattice constant radius. Knowing the exchange coupling parameters of the system as evaluated by mapping the system onto the classical Heisenberg Hamiltonian, one can rather straightforwardly calculate the Curie temperature by means of MFA~\cite{Anderson1963,Rancourt1993,Dang1996,Sasioglu2004,Takahashi2013,Wipf2013}. Note, however, that this simplified approach totally neglects fluctuations.

To evaluate the Curie temperature of a specific alloy beyond MFA, we also conduct the Monte Carlo simulations of the net magnetization as a function of temperature using the \texttt{VAMPIRE} atomistic spin dynamic program~\cite{Evans2020,Vampire}, where the Curie temperature is extracted from the Curie-Bloch equation in the classical limit~\cite{Evans2015}. The system is emulated by a cube with side length 10 nm and periodic boundary conditions, provided that some fraction of host atoms Ni are replaced by alloy atoms Fe. Similar to MFA calculations, the unit cell parameters are adjusted to experimental values (Table~\ref{tab:latmom}). For each temperature, we perform the spin thermalization of the system using 5000 Monte Carlo steps and subsequently measure its thermal equilibrium magnetization by averaging of the following 5000 steps. In our simulations, we emulate the classical Heisenberg model with exchange interaction between atoms up to the third coordination shell extracted from {\it ab initio} results, and no magneto-crystalline anisotropy is present.

Both MFA and Monte Carlo simulations do not respect the transverse and longitudinal spin fluctuations that are of practical importance for the correct description of magnetism in Fe and Ni~\cite{Ruban2007,MD12}. In Ref.~\cite{MD12}, the longitudinal spin fluctuations (LSF) are included at the level of spin dynamics simulations, revealing a better estimate of the Curie temperature for Fe. The role of LSF is even more pronounced for Ni as demonstrated in Ref.~\cite{Ruban2007}, where Monte Carlo simulations with the effective classical Hamiltonian incorporate spin fluctuations. The account of LSF results in a better agreement between the predicted Curie temperature for Ni and the experimental results in comparison to rigid spin calculations, such as MFA. However, the methodology developed in Refs.~\cite{Ruban2007,MD12} has its own limitations stemming from a disregard of the quantum character of spin fluctuations. For instance, the calculated magnetization vs. temperature curve in the spin dynamics simulations with LSF~\cite{MD12} clearly manifests that the reasonable agreement of the Curie temperature 
is achieved by a too fast decrease of magnetization at low temperatures. 

In this paper, we examine iron-nickel alloys in the framework of spin-fluctuation theory~\cite{Moriya85} that takes into account the quantum character of spin fluctuations using the DSFT approach. The detailed description of DSFT formalism is given in Ref.~\cite{MR18}, and here we only briefly outline its key points. The starting point of the DSFT is the multiband Hubbard model with on-site repulsion between localized $d$ electrons. The pair interaction between electrons is replaced by the interaction of each electron with the fluctuating field by means of the Hubbard-Stratonovich transformation. Following that, the magnetic properties of the system are evaluated within the functional integral method by averaging over all possible configurations of the field. For each temperature, the probability density of this fluctuating field is calculated self-consistently in the Gaussian approximation. The initial data for the DSFT calculations are the averaged spin magnetic moment of the system $m_0$ at zero temperature and the non-magnetic DOS obtained from  first principles.

We construct the nonmagnetic DOS of the Fe$_{x}$Ni$_{1-x}$ alloy from the spin-resolved DOS using the method elaborated in Ref.~\cite{Res07}. First, we distinguish only the $d$ band contribution to the spin-resolved DOS of Fe and Ni sites ($\nu_{\sigma}^{\mathrm{Fe,Ni}}$) and calculate the total DOS as 
\begin{equation*}
  \nu_{\sigma}^{\mathrm{tot}}(E) 
  = x\nu_{\sigma}^{\mathrm{Fe}}(E) + (1-x)\nu_{\sigma}^{\mathrm{Ni}}(E),
\end{equation*}
where the subscript $\sigma$ stands for the spin degree of freedom. Next, we shift the spin-resolved DOSs to each other by $V_0$ and sum them,
\begin{equation*}
  \nu(E) 
  = \nu_{\uparrow}^{\mathrm{tot}}(E - V_0)  
  + \nu_{\downarrow}^{\mathrm{tot}}(E + V_0). 
\end{equation*}
The energy shift $V_0$ and nonmagnetic Fermi level $E_F$ are found from the system of two nonlinear equations,
\begin{equation}\label{nmdos-eq}
    n_{\uparrow} - n_{\downarrow} = 0,
    \qquad
    n_{\uparrow} +n_{\downarrow} - n_{\mathrm{e}} = 0.
\end{equation}
Here, the first equation describes the condition of zero magnetization and the second one ensures that the number of electrons is conserved. In Eqs.~\eqref{nmdos-eq},
\begin{equation*}
  n_{\sigma} 
  = \int_{0}^{E_F} 
  \nu^{\mathrm{tot}}_{\sigma}(E - \sigma V_0) 
  \, d E
\end{equation*}
is the number of electrons with spin projection $\sigma$ per atom and $n_{\mathrm{e}}$ is the total number of electrons per atom. Note that the constructed nonmagnetic DOS is introduced for the problem simplification, meaning that the Fe and Ni atoms are replaced by ``effective medium'' atoms with an average $d$ band. However, as pointed out in Refs.~\cite{Res07,MR18}, it should not be crucial because of the integral dependence of the DSFT equations on the electronic energy structure. Finally, the obtained DOS is normalized to 10 (the number of $d$ states per atom) and slightly smoothed by a convolution with the Lorentzian function with the half width $\Gamma = 0.001 \, W$, where $W$ is the bandwidth equal to 10.74~eV for all alloy compositions $x$ in our calculations.

In the DSFT, the temperature dependence of magnetic characteristics is calculated by the numerical continuation method~\cite{PMR20}. At each temperature, one needs to solve the set of nonlinear equations that consists of four equations with respect to scalar variables: the chemical potential $\mu$, mean field $\bar V^z$, transverse $\zeta^x$, and longitudinal $\zeta^z$ single-site spin fluctuations, and two equations with respect to complex functions: the coherent potential and single-site Green's function. We performed these calculations using the \texttt{MAGPROP} program suite \cite{RPM18}. Note that results presented below are carried out in the renormalized Gaussian approximation of the DSFT that allows one to get a better agreement with experiment and eliminate any possible hysteresis behavior in the temperature dependence calculated in the Gaussian approximation (for details and application to the Invar alloy Fe$_{0.65}$Ni$_{0.35}$, see Refs.~\cite{MRG11,Reser2009}).

\section{Electronic properties}

To investigate the finite-temperature properties of Fe-Ni alloys, we first examine their electronic structure depending on the alloy composition. The Fermi energy given relative to the muffin-tin zero $E_F$ and spin magnetic moments of Fe and Ni atoms, $m^\mathrm{Fe,Ni}$, as well as the averaged moment of the system $m_0$ as computed by virtue of the KKR-CPA approach are depicted in Table~\ref{tab:latmom}. One can clearly notice that the spin magnetic moment of Fe decreases with $x$, whereas that of Ni is almost independent of alloy composition. As a result, the averaged spin magnetic moment calculated as a combination of atomic magnetic moments per unit cell increases linearly with $x$. Our numerical findings on magnetic moments are in reasonable agreement with previously reported theoretical~\cite{Fu2019, Weinberger2001,James1999,MSL02} and experimental~\cite{Swartzendruber1991,Crangle1963} data also shown in Table~\ref{tab:latmom}.

\begin{table}[ht!]
\small
\caption{\label{tab:latmom} Composition dependence of the Fermi energy $E_F$ and the element-resolved $m^\mathrm{Ni,Fe}$ and total $m_0$ spin magnetic moments in Fe$_x$Ni$_{1-x}$ alloys calculated using the KKR-CPA formalism with the lattice parameter $a$ as provided in Ref.~\cite{Glaubitz2011} as well as the experimentally available averaged magnetic moments $m_0^\mathrm{exp}$. All the lattice parameters are given in \AA\, and the energies are in eV, whereas all the magnetic moments are in Bohr magneton units.}
\begin{tabular*}{0.48\textwidth}{@{\extracolsep{\fill}}lllllll}
\hline
 $x$ &   $a$ &  $E_F$ & $m^\mathrm{Ni}$ &  $m^\mathrm{Fe}$ & $m_0$ & $m_0^\mathrm{exp}$ \\ 
\hline
0.1 & 3.536 & 9.485 & 0.67 & 2.63 & 0.86 & 0.81\cite{Crangle1963}\\
0.2 & 3.548 & 9.501 & 0.67 & 2.61 & 1.05 & 1.06 \cite{Swartzendruber1991}, 1.03 \cite{Crangle1963}\\ 
0.3 & 3.559 & 9.414 & 0.68 & 2.59 & 1.25 & 1.26 \cite{Crangle1963}\\
0.4 & 3.574 & 9.493 & 0.70 & 2.58 & 1.45 & 1.50\cite{Swartzendruber1991,Crangle1963}\\ 
0.5 & 3.587 & 9.487 & 0.70 & 2.55 & 1.62 & 1.69 \cite{Swartzendruber1991}, 1.65 \cite{Crangle1963}\\
0.6 & 3.596 & 9.512 & 0.69 & 2.51 & 1.78 & 1.80\cite{Swartzendruber1991}, 1.78 \cite{Crangle1963}\\
\hline
\end{tabular*}
\end{table}

\begin{figure}[ht!]
\centering
\includegraphics[scale=0.48]{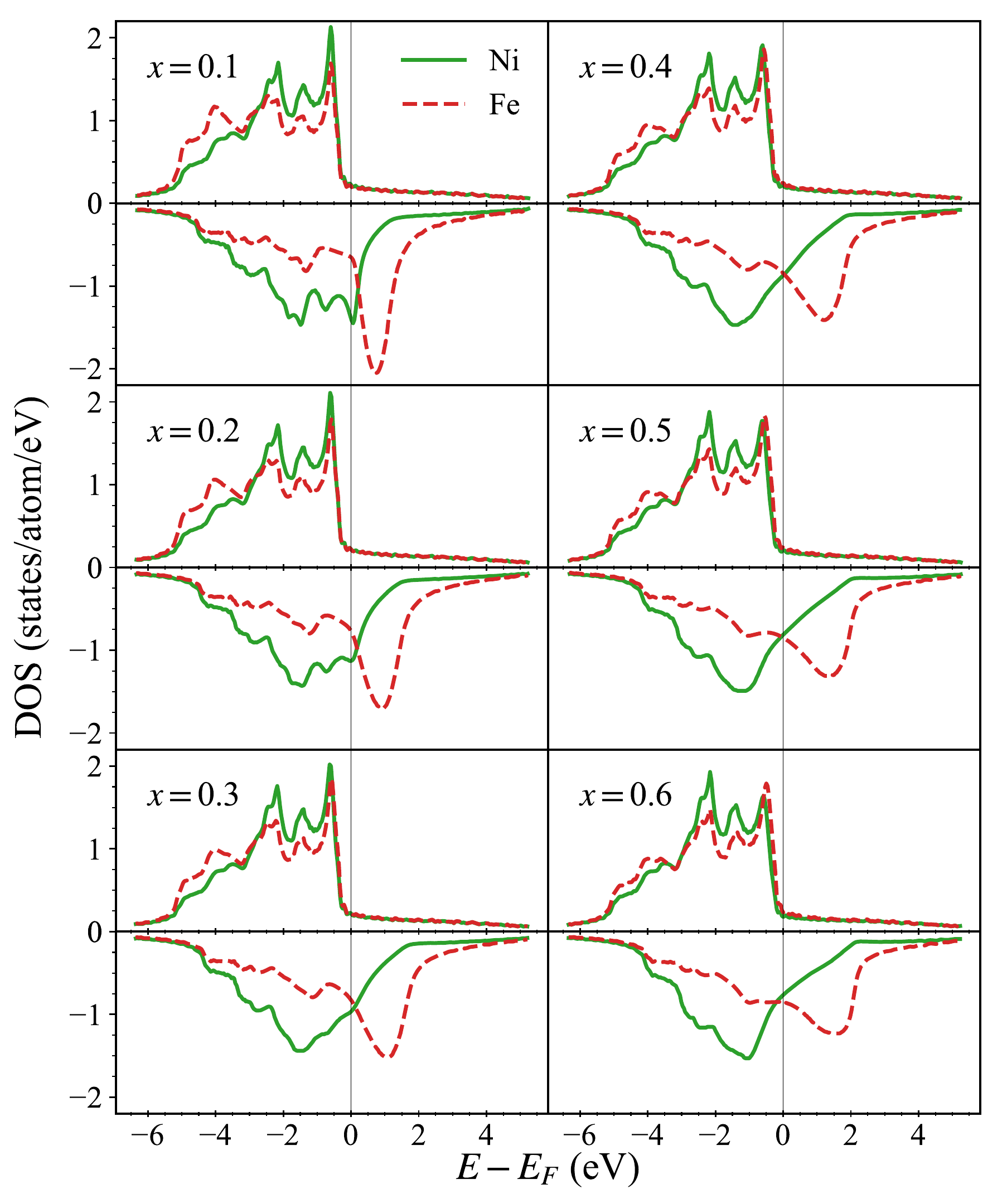}
\caption{Spin-resolved density of states of Fe (red dashed lines) and Ni (green solid lines) sites in Fe$_x$Ni$_{1-x}$ alloy calculated within the scalar-relativistic KKR-CPA scheme. The top half of each panel refers to the contribution from majority spin (spin up) and the bottom half to minority spin (spin down). The vertical gray line marks the zero energy positioned at the Fermi level, $E_F$.}
\label{fig:dos}
\end{figure}

\begin{figure}
\centering
\includegraphics[width=1.0\linewidth]{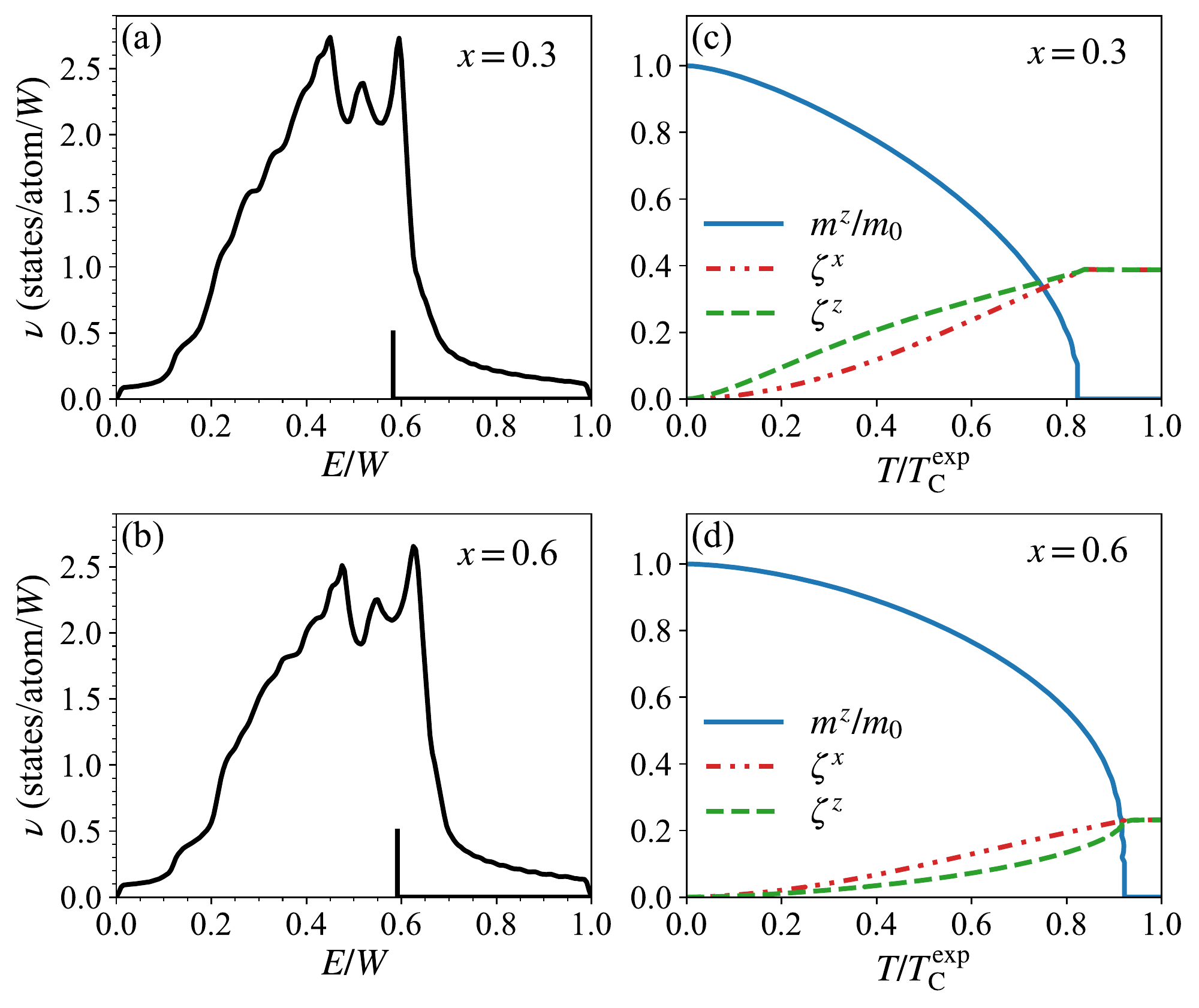}
\caption{Nonmagnetic DOS $\nu(E)$ of the $d$ band of Fe$_{x}$Ni$_{1-x}$ for $x=0.3$ (a) and 0.6 (b), respectively. The energy $E$ is measured in the units of the bandwidth $W = 10.74$ eV, and the vertical line is positioned at the Fermi level. The magnetization $m^z/m_0$, transverse $\zeta^x$, and longitudinal $\zeta^z$ spin fluctuations normalized by the squared mean exchange field $\bar{V}_z^2$ at $T=0$ of Fe$_{x}$Ni$_{1-x}$ for $x=0.3$ (c) and 0.6 (d), respectively, in the renormalized Gaussian approximation of the DSFT.}
\label{fig:DSFT}
\end{figure}

In Fig.~\ref{fig:dos}, we present the calculated spin-resolved DOS curves for different alloy composition. As one can notice from Table~\ref{tab:latmom}, the position of the Fermi energy remains almost unchanged with the increase of Fe concentration. A close inspection of the majority spin (spin up) DOS of Fe and Ni reveal almost no response to the composition change, meaning that they remain unperturbed under constituent variation. In contrast, the maximum attributed to minority spin (spin down) DOS of Fe decreases and shifts to the higher energy level with $x$. Note that the calculated density of states is almost the same as the material spin-resolved DOS found at minimal sufficient angular momentum cutoff value ($l_{max}=2$). The calculated spin-polarized densities are utilized for constructing the nonmagnetic DOS $\nu(E)$ of Fe$_{x}$Ni$_{1-x}$ that is used as initial data for the DSFT calculations. Following the discussed routine, we compute $\nu(E)$ for each alloy composition and present the nonmagnetic DOS for the case $x=0.3$ and 0.6 at Figs.~\ref{fig:DSFT}(a) and \ref{fig:DSFT}(b).

\begin{figure}[h!]
\centering
\includegraphics[scale=0.5]{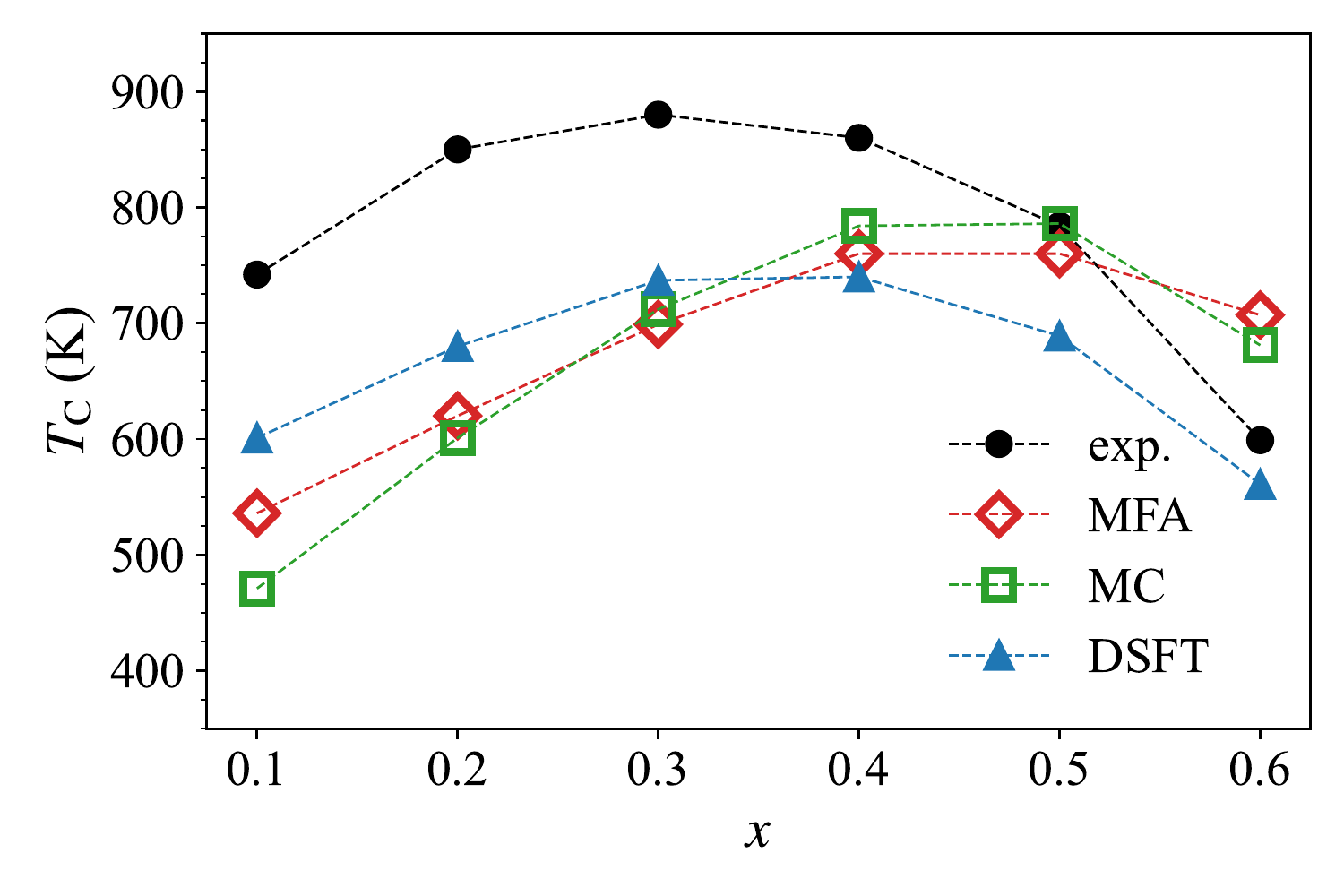}
\caption{Curie temperature $T_{\mathrm{C}}$ of random Fe$_x$Ni$_{1-x}$ alloys calculated using MFA (red diamonds), Monte Carlo simulations (green squares), and DSFT (blue triangles) as a function of alloy composition. The experimental values~\cite{Swartzendruber1991,Kudrnovsky2008} are marked by black circles.}
\label{fig:Curie}
\end{figure}

We provide the exchange coupling strength depending on alloy composition in Table~\ref{tab:exchange}. In $0.1\leq x\leq0.4$ concentration range, the increase of iron leads to a significant increase of Fe-Fe interaction that subsequently decreases for $x>0.4$. One can expect that this enhancement at around 30--40\% of iron concentration might lead to the rise of the Curie temperature and will be discussed in the following. The exchange coupling strength between nickel atoms is about one order of magnitude smaller than that between iron atoms, while the interspecies exchange between nickel and iron continuously decreases with Fe excess favoring ferromagnetic ordering. 
\begin{table}[]
\centering
\caption{\label{tab:exchange} Dependence of the exchange coupling strength for the 1st, 2nd, and 3rd coordination shell in random alloys of Fe$_x$Ni$_{1-x}$ given in meV as varied with alloy composition.}
\begin{tabular}{c|ccc|ccc|ccc}
\hline
    & \multicolumn{3}{c|}{$J_{\rm Ni-Ni}$} & \multicolumn{3}{c|}{$J_{\rm Ni-Fe}$} & \multicolumn{3}{c}{$J_{\rm Fe-Fe}$} \\
$x$ & 1st        & 2nd         & 3rd       & 1st        & 2nd        & 3rd        & 1st        & 2nd       & 3rd        \\ \hline
0.1 & 2.70        & -0.12       & 0.37      & 9.72       & 1.03       & 1.49       & 8.95       & 6.63      & 1.36       \\
0.2 & 2.44       & -0.20       & 0.35      & 9.46       & 0.93       & 1.35       & 9.91       & 7.16      & 1.11       \\
0.3 & 2.29       & -0.23       & 0.35      & 9.29       & 0.93       & 1.22       & 10.8       & 7.2       & 0.83       \\
0.4 & 2.24       & -0.23       & 0.35      & 9.16       & 0.89       & 1.09       & 11.2       & 6.94      & 0.44       \\
0.5 & 2.16       & -0.22       & 0.36      & 8.78       & 0.84       & 0.93       & 10.4       & 6.67      & -0.20      \\
0.6 & 2.09       & -0.23       & 0.36      & 8.28       & 0.77       & 0.73       & 8.99       & 6.88      & -1.23      \\ \hline
\end{tabular}
\label{tab:exchange}
\end{table} 

\section{Curie temperature}

According to experimental results, the Curie temperature of disordered Fe$_x$Ni$_{1-x}$ alloys exhibits a nonmonotonous behavior upon the increase of iron concentration with reaching maximum value at $x\simeq0.3$~\cite{Swartzendruber1991,Kudrnovsky2008}. The Curie temperatures of random binaries obtained by MFA, Monte Carlo, and DSFT methods are shown in Fig.~\ref{fig:Curie}. Clearly, all the methods tend to underestimate the Curie temperature as compared to the experimental data. However, it is interesting to note that the use of MFA and Monte Carlo results in the Curie temperature reaching its maximum at $x\simeq 0.5$, which leads to the conclusion of insufficiency of these methods for qualitative description of alloy magnetic properties. Particularly, the correct behavior can not be captured in the Monte Carlo framework, despite the maximum of Fe-Fe exchange coupling strength at about 30--40\% of iron concentration (see Table~\ref{tab:exchange}) included into the system magnetization simulations. For specific alloy compositions, to get the temperature close to experimental value using Monte Carlo simulations, the exchange parameter should be sufficiently increased~\cite{Evans2014,Evans2015}.

Progressing from MFA and Monte Carlo simulations to the DSFT, we notice that the latter keeps track of the temperature trend revealing the importance of both longitudinal and transverse spin fluctuations for describing magnetic properties at high temperatures. To check their impact with respect to the temperature, at Figs.~\ref{fig:DSFT}(c) and \ref{fig:DSFT}(d) we show the basic magnetic characteristics of Fe$_{x}$Ni$_{1-x}$ calculated for $x=0.3$ and 0.6. Notice that for small iron concentration ($x=0.3$) the longitudinal spin fluctuations $\zeta^z$ dominate over the transverse ones $\zeta^x$, while the situation is opposite when $x=0.6$, which is in line with the previous analysis of pure Fe and Ni systems, as well as the Invar alloy~\cite{Res07,MRG11,MR18}.

For selected iron concentrations, the Curie temperature of Fe$_x$Ni$_{1-x}$ alloys has been also estimated using Monte Carlo simulations with the effective Heisenberg Hamiltonian including LSF as reported in Refs.~\cite{Ruban2017,Pan2017}. For permalloy, i.e., $x=0.2$, $T_{\rm C}$ varies from 482~K to 572~K depending on the utilized approach~\cite{Pan2017}, whereas for $x=0.3$ and $0.5$, the reported $T_{\rm C}$ values are $800$ K and $650$ K~\cite{Ruban2017}, respectively. However, the systematic calculation of $T_{\rm C}$ as a function of $x$ in these simulations is absent. The systematic study of the $T_{\rm C}$ dependence on $x$ for Fe$_x$Ni$_{1-x}$ alloys has been addressed in Ref.~\cite{Kudrnovsky2008}. The best agreement has been obtained by means of the renormalized random phase approximation (rRPA), where the random alloys are treated as crystals with the effective exchange interactions~\cite{Bruno03}. In comparison to the DSFT, the obtained rRPA results demonstrate better quantitative agreement, but it does not catch the qualitative trend of $T_{\rm C}$ properly similar to our results found using MFA and MC. Indeed, the rRPA gives the monotonous increase of $T_{\rm C}$ up to 864 K at $x \simeq 0.5$ in contrast to maximum $T_{\rm C} = 880$ K at $x \simeq 0.3$ observed in experiments. For $x > 0.5$, the rRPA overestimates the Curie temperature. This is attributed to the fact that the rRPA does not respect spin fluctuations.

\section{Conclusions}

In summary, we have systematically investigated the electronic and magnetic properties of disordered Fe$_x$Ni$_{1-x}$ alloys in the concentration range from $x=0.1$ to 0.6 of iron using the results of first-principles calculations. Within the KKR-CPA approach, we have calculated the element-resolved and averaged magnetic moments of selected alloy compositions and found their agreement with experimental values. We have noticed the strong impact of iron concentration on spin-resolved density of states of specific alloys and exchange coupling strength. Applying the DSFT, we have estimated the Curie temperature of random iron-nickel binaries as varied upon increasing iron concentration. We benchmark the DSFT results against the available experimental data on the Curie temperature: as opposed to MFA and Monte Carlo simulations, the DSFT demonstrates good agreement and unambiguously reveals the leading role of spin fluctuations. Moreover, comparing our results to the previous systematic calculations of the Curie temperature by the rRPA, we show that the DSFT improves the rRPA results by correctly reproducing the qualitative behaviour of $T_{\rm C}$ upon iron concentration increase. \\

\acknowledgements
The work of A.A.P. was supported by the Russian Science Foundation Project No.~20-72-00044 (first-principles calculations, atomistic spin dynamics simulations, and analysis of the results). The authors express their gratitude to the group of Professor H. Ebert for providing the SPR-KKR package \cite{SPRKKR}, the group of Professor R. F. L. Evans for sharing the \texttt{VAMPIRE} software package \cite{Vampire}, and the group of Dr. N. B. Melnikov for sharing the \texttt{MAGPROP} software package \cite{RPM18}. The authors acknowledge the use of the ``Zhores'' supercomputer \cite{Zhores} for obtaining the results presented in this paper. 

\bibliographystyle{apsrev4-1}
\bibliography{main.bbl}

\providecommand{\noopsort}[1]{}\providecommand{\singleletter}[1]{#1}%
\begin{thebibliography}{95}%
\makeatletter
\providecommand \@ifxundefined [1]{%
 \@ifx{#1\undefined}
}%
\providecommand \@ifnum [1]{%
 \ifnum #1\expandafter \@firstoftwo
 \else \expandafter \@secondoftwo
 \fi
}%
\providecommand \@ifx [1]{%
 \ifx #1\expandafter \@firstoftwo
 \else \expandafter \@secondoftwo
 \fi
}%
\providecommand \natexlab [1]{#1}%
\providecommand \enquote  [1]{``#1''}%
\providecommand \bibnamefont  [1]{#1}%
\providecommand \bibfnamefont [1]{#1}%
\providecommand \citenamefont [1]{#1}%
\providecommand \href@noop [0]{\@secondoftwo}%
\providecommand \href [0]{\begingroup \@sanitize@url \@href}%
\providecommand \@href[1]{\@@startlink{#1}\@@href}%
\providecommand \@@href[1]{\endgroup#1\@@endlink}%
\providecommand \@sanitize@url [0]{\catcode `\\12\catcode `\$12\catcode
  `\&12\catcode `\#12\catcode `\^12\catcode `\_12\catcode `\%12\relax}%
\providecommand \@@startlink[1]{}%
\providecommand \@@endlink[0]{}%
\providecommand \url  [0]{\begingroup\@sanitize@url \@url }%
\providecommand \@url [1]{\endgroup\@href {#1}{\urlprefix }}%
\providecommand \urlprefix  [0]{URL }%
\providecommand \Eprint [0]{\href }%
\providecommand \doibase [0]{http://dx.doi.org/}%
\providecommand \selectlanguage [0]{\@gobble}%
\providecommand \bibinfo  [0]{\@secondoftwo}%
\providecommand \bibfield  [0]{\@secondoftwo}%
\providecommand \translation [1]{[#1]}%
\providecommand \BibitemOpen [0]{}%
\providecommand \bibitemStop [0]{}%
\providecommand \bibitemNoStop [0]{.\EOS\space}%
\providecommand \EOS [0]{\spacefactor3000\relax}%
\providecommand \BibitemShut  [1]{\csname bibitem#1\endcsname}%
\let\auto@bib@innerbib\@empty
\bibitem [{\citenamefont {Swartzendruber}\ \emph {et~al.}(1991)\citenamefont
  {Swartzendruber}, \citenamefont {Itkin},\ and\ \citenamefont
  {Alcock}}]{Swartzendruber1991}%
  \BibitemOpen
  \bibfield  {author} {\bibinfo {author} {\bibfnamefont {L.~J.}\ \bibnamefont
  {Swartzendruber}}, \bibinfo {author} {\bibfnamefont {V.~P.}\ \bibnamefont
  {Itkin}}, \ and\ \bibinfo {author} {\bibfnamefont {C.~B.}\ \bibnamefont
  {Alcock}},\ }\href@noop {} {\bibfield  {journal} {\bibinfo  {journal} {J.
  Phase Equilibria}\ }\textbf {\bibinfo {volume} {12}},\ \bibinfo {pages} {288}
  (\bibinfo {year} {1991})}\BibitemShut {NoStop}%
\bibitem [{\citenamefont {Mao}\ \emph {et~al.}(2006)\citenamefont {Mao},
  \citenamefont {Campbell}, \citenamefont {Heinz},\ and\ \citenamefont
  {Shen}}]{Mao2006}%
  \BibitemOpen
  \bibfield  {author} {\bibinfo {author} {\bibfnamefont {W.~L.}\ \bibnamefont
  {Mao}}, \bibinfo {author} {\bibfnamefont {A.~J.}\ \bibnamefont {Campbell}},
  \bibinfo {author} {\bibfnamefont {D.~L.}\ \bibnamefont {Heinz}}, \ and\
  \bibinfo {author} {\bibfnamefont {G.}~\bibnamefont {Shen}},\ }\href {\doibase
  https://doi.org/10.1016/j.pepi.2005.11.002} {\bibfield  {journal} {\bibinfo
  {journal} {Phys. Earth Planet. Inter.}\ }\textbf {\bibinfo {volume} {155}},\
  \bibinfo {pages} {146} (\bibinfo {year} {2006})}\BibitemShut {NoStop}%
\bibitem [{\citenamefont {Cacciamani}\ \emph {et~al.}(2010)\citenamefont
  {Cacciamani}, \citenamefont {Dinsdale}, \citenamefont {Palumbo},\ and\
  \citenamefont {Pasturel}}]{Cacciamani2010}%
  \BibitemOpen
  \bibfield  {author} {\bibinfo {author} {\bibfnamefont {G.}~\bibnamefont
  {Cacciamani}}, \bibinfo {author} {\bibfnamefont {A.}~\bibnamefont
  {Dinsdale}}, \bibinfo {author} {\bibfnamefont {M.}~\bibnamefont {Palumbo}}, \
  and\ \bibinfo {author} {\bibfnamefont {A.}~\bibnamefont {Pasturel}},\ }\href
  {\doibase https://doi.org/10.1016/j.intermet.2010.02.026} {\bibfield
  {journal} {\bibinfo  {journal} {Intermetallics}\ }\textbf {\bibinfo {volume}
  {18}},\ \bibinfo {pages} {1148} (\bibinfo {year} {2010})}\BibitemShut
  {NoStop}%
\bibitem [{\citenamefont {Glaubitz}\ \emph {et~al.}(2011)\citenamefont
  {Glaubitz}, \citenamefont {Buschhorn}, \citenamefont {Br{\"u}ssing},
  \citenamefont {Abrudan},\ and\ \citenamefont {Zabel}}]{Glaubitz2011}%
  \BibitemOpen
  \bibfield  {author} {\bibinfo {author} {\bibfnamefont {B.}~\bibnamefont
  {Glaubitz}}, \bibinfo {author} {\bibfnamefont {S.}~\bibnamefont {Buschhorn}},
  \bibinfo {author} {\bibfnamefont {F.}~\bibnamefont {Br{\"u}ssing}}, \bibinfo
  {author} {\bibfnamefont {R.}~\bibnamefont {Abrudan}}, \ and\ \bibinfo
  {author} {\bibfnamefont {H.}~\bibnamefont {Zabel}},\ }\href@noop {}
  {\bibfield  {journal} {\bibinfo  {journal} {J. Phys. Condens. Matter}\
  }\textbf {\bibinfo {volume} {23}},\ \bibinfo {pages} {254210} (\bibinfo
  {year} {2011})}\BibitemShut {NoStop}%
\bibitem [{\citenamefont {Xiong}\ \emph {et~al.}(2011)\citenamefont {Xiong},
  \citenamefont {Zhang}, \citenamefont {Vitos},\ and\ \citenamefont
  {Selleby}}]{Xiong2011}%
  \BibitemOpen
  \bibfield  {author} {\bibinfo {author} {\bibfnamefont {W.}~\bibnamefont
  {Xiong}}, \bibinfo {author} {\bibfnamefont {H.}~\bibnamefont {Zhang}},
  \bibinfo {author} {\bibfnamefont {L.}~\bibnamefont {Vitos}}, \ and\ \bibinfo
  {author} {\bibfnamefont {M.}~\bibnamefont {Selleby}},\ }\href {\doibase
  https://doi.org/10.1016/j.actamat.2010.09.055} {\bibfield  {journal}
  {\bibinfo  {journal} {Acta Mater.}\ }\textbf {\bibinfo {volume} {59}},\
  \bibinfo {pages} {521} (\bibinfo {year} {2011})}\BibitemShut {NoStop}%
\bibitem [{\citenamefont {Kakehashi}(1981)}]{Kakehashi1981}%
  \BibitemOpen
  \bibfield  {author} {\bibinfo {author} {\bibfnamefont {Y.}~\bibnamefont
  {Kakehashi}},\ }\href@noop {} {\bibfield  {journal} {\bibinfo  {journal} {J.
  Phys. Soc. Jpn.}\ }\textbf {\bibinfo {volume} {50}},\ \bibinfo {pages} {1925}
  (\bibinfo {year} {1981})}\BibitemShut {NoStop}%
\bibitem [{\citenamefont {Moruzzi}(1990)}]{Moruzzi1990}%
  \BibitemOpen
  \bibfield  {author} {\bibinfo {author} {\bibfnamefont {V.~L.}\ \bibnamefont
  {Moruzzi}},\ }\href {\doibase https://doi.org/10.1016/0921-4526(89)90112-9}
  {\bibfield  {journal} {\bibinfo  {journal} {Physica B Condens. Matter}\
  }\textbf {\bibinfo {volume} {161}},\ \bibinfo {pages} {99} (\bibinfo {year}
  {1990})}\BibitemShut {NoStop}%
\bibitem [{\citenamefont {Wasserman}(1990)}]{Wasserman1990}%
  \BibitemOpen
  \bibfield  {author} {\bibinfo {author} {\bibfnamefont {E.~F.}\ \bibnamefont
  {Wasserman}}\ }(\bibinfo  {publisher} {Elsevier},\ \bibinfo {address}
  {Amsterdam},\ \bibinfo {year} {1990})\ pp.\ \bibinfo {pages}
  {237--322}\BibitemShut {NoStop}%
\bibitem [{\citenamefont {Abrikosov}\ \emph {et~al.}(1995)\citenamefont
  {Abrikosov}, \citenamefont {Eriksson}, \citenamefont {S\"oderlind},
  \citenamefont {Skriver},\ and\ \citenamefont {Johansson}}]{Abrikosov1995}%
  \BibitemOpen
  \bibfield  {author} {\bibinfo {author} {\bibfnamefont {I.~A.}\ \bibnamefont
  {Abrikosov}}, \bibinfo {author} {\bibfnamefont {O.}~\bibnamefont {Eriksson}},
  \bibinfo {author} {\bibfnamefont {P.}~\bibnamefont {S\"oderlind}}, \bibinfo
  {author} {\bibfnamefont {H.~L.}\ \bibnamefont {Skriver}}, \ and\ \bibinfo
  {author} {\bibfnamefont {B.}~\bibnamefont {Johansson}},\ }\href {\doibase
  10.1103/PhysRevB.51.1058} {\bibfield  {journal} {\bibinfo  {journal} {Phys.
  Rev. B}\ }\textbf {\bibinfo {volume} {51}},\ \bibinfo {pages} {1058}
  (\bibinfo {year} {1995})}\BibitemShut {NoStop}%
\bibitem [{\citenamefont {Schr\"oter}\ \emph {et~al.}(1995)\citenamefont
  {Schr\"oter}, \citenamefont {Ebert}, \citenamefont {Akai}, \citenamefont
  {Entel}, \citenamefont {Hoffmann},\ and\ \citenamefont
  {Reddy}}]{Schroter1995}%
  \BibitemOpen
  \bibfield  {author} {\bibinfo {author} {\bibfnamefont {M.}~\bibnamefont
  {Schr\"oter}}, \bibinfo {author} {\bibfnamefont {H.}~\bibnamefont {Ebert}},
  \bibinfo {author} {\bibfnamefont {H.}~\bibnamefont {Akai}}, \bibinfo {author}
  {\bibfnamefont {P.}~\bibnamefont {Entel}}, \bibinfo {author} {\bibfnamefont
  {E.}~\bibnamefont {Hoffmann}}, \ and\ \bibinfo {author} {\bibfnamefont
  {G.~G.}\ \bibnamefont {Reddy}},\ }\href {\doibase 10.1103/PhysRevB.52.188}
  {\bibfield  {journal} {\bibinfo  {journal} {Phys. Rev. B}\ }\textbf {\bibinfo
  {volume} {52}},\ \bibinfo {pages} {188} (\bibinfo {year} {1995})}\BibitemShut
  {NoStop}%
\bibitem [{\citenamefont {Wassermann}(1991)}]{Wassermann1991}%
  \BibitemOpen
  \bibfield  {author} {\bibinfo {author} {\bibfnamefont {E.~F.}\ \bibnamefont
  {Wassermann}},\ }\href@noop {} {\bibfield  {journal} {\bibinfo  {journal} {J.
  Magn. Magn. Mater.}\ }\textbf {\bibinfo {volume} {100}},\ \bibinfo {pages}
  {346} (\bibinfo {year} {1991})}\BibitemShut {NoStop}%
\bibitem [{\citenamefont {van Schilfgaarde}\ \emph {et~al.}(1999)\citenamefont
  {van Schilfgaarde}, \citenamefont {Abrikosov},\ and\ \citenamefont
  {Johansson}}]{Van1999}%
  \BibitemOpen
  \bibfield  {author} {\bibinfo {author} {\bibfnamefont {M.}~\bibnamefont {van
  Schilfgaarde}}, \bibinfo {author} {\bibfnamefont {I.~A.}\ \bibnamefont
  {Abrikosov}}, \ and\ \bibinfo {author} {\bibfnamefont {B.}~\bibnamefont
  {Johansson}},\ }\href@noop {} {\bibfield  {journal} {\bibinfo  {journal}
  {Nature}\ }\textbf {\bibinfo {volume} {400}},\ \bibinfo {pages} {46}
  (\bibinfo {year} {1999})}\BibitemShut {NoStop}%
\bibitem [{\citenamefont {Khmelevskyi}\ and\ \citenamefont
  {Mohn}(2004)}]{Khmelevskyi2004}%
  \BibitemOpen
  \bibfield  {author} {\bibinfo {author} {\bibfnamefont {S.}~\bibnamefont
  {Khmelevskyi}}\ and\ \bibinfo {author} {\bibfnamefont {P.}~\bibnamefont
  {Mohn}},\ }\href {\doibase 10.1103/PhysRevB.69.140404} {\bibfield  {journal}
  {\bibinfo  {journal} {Phys. Rev. B}\ }\textbf {\bibinfo {volume} {69}},\
  \bibinfo {pages} {140404} (\bibinfo {year} {2004})}\BibitemShut {NoStop}%
\bibitem [{\citenamefont {Asano}(1969)}]{Asano1969}%
  \BibitemOpen
  \bibfield  {author} {\bibinfo {author} {\bibfnamefont {H.}~\bibnamefont
  {Asano}},\ }\href@noop {} {\bibfield  {journal} {\bibinfo  {journal} {J.
  Phys. Soc. Jpn.}\ }\textbf {\bibinfo {volume} {27}},\ \bibinfo {pages} {542}
  (\bibinfo {year} {1969})}\BibitemShut {NoStop}%
\bibitem [{\citenamefont {Window}(1973)}]{Window1973}%
  \BibitemOpen
  \bibfield  {author} {\bibinfo {author} {\bibfnamefont {B.}~\bibnamefont
  {Window}},\ }\href@noop {} {\bibfield  {journal} {\bibinfo  {journal} {J.
  Appl. Phys.}\ }\textbf {\bibinfo {volume} {44}},\ \bibinfo {pages} {2853}
  (\bibinfo {year} {1973})}\BibitemShut {NoStop}%
\bibitem [{\citenamefont {Sedov}\ and\ \citenamefont
  {Tsigel'nik}(1998)}]{Sedov1998}%
  \BibitemOpen
  \bibfield  {author} {\bibinfo {author} {\bibfnamefont {V.~L.}\ \bibnamefont
  {Sedov}}\ and\ \bibinfo {author} {\bibfnamefont {O.~A.}\ \bibnamefont
  {Tsigel'nik}},\ }\href {\doibase
  https://doi.org/10.1016/S0304-8853(97)01076-7} {\bibfield  {journal}
  {\bibinfo  {journal} {J. Magn. Magn. Mater.}\ }\textbf {\bibinfo {volume}
  {183}},\ \bibinfo {pages} {117} (\bibinfo {year} {1998})}\BibitemShut
  {NoStop}%
\bibitem [{\citenamefont {Bozorth}(1953)}]{Bozorth1953}%
  \BibitemOpen
  \bibfield  {author} {\bibinfo {author} {\bibfnamefont {R.~M.}\ \bibnamefont
  {Bozorth}},\ }\href {\doibase 10.1103/RevModPhys.25.42} {\bibfield  {journal}
  {\bibinfo  {journal} {Rev. Mod. Phys.}\ }\textbf {\bibinfo {volume} {25}},\
  \bibinfo {pages} {42} (\bibinfo {year} {1953})}\BibitemShut {NoStop}%
\bibitem [{\citenamefont {Ferguson}(1958)}]{Ferguson1958}%
  \BibitemOpen
  \bibfield  {author} {\bibinfo {author} {\bibfnamefont {E.~T.}\ \bibnamefont
  {Ferguson}},\ }\href@noop {} {\bibfield  {journal} {\bibinfo  {journal} {J.
  Appl. Phys.}\ }\textbf {\bibinfo {volume} {29}},\ \bibinfo {pages} {252}
  (\bibinfo {year} {1958})}\BibitemShut {NoStop}%
\bibitem [{\citenamefont {Pfeifer}\ and\ \citenamefont
  {Radeloff}(1980)}]{Pfeifer1980}%
  \BibitemOpen
  \bibfield  {author} {\bibinfo {author} {\bibfnamefont {F.}~\bibnamefont
  {Pfeifer}}\ and\ \bibinfo {author} {\bibfnamefont {C.}~\bibnamefont
  {Radeloff}},\ }\href {\doibase https://doi.org/10.1016/0304-8853(80)90592-2}
  {\bibfield  {journal} {\bibinfo  {journal} {J. Magn. Magn. Mater.}\ }\textbf
  {\bibinfo {volume} {19}},\ \bibinfo {pages} {190} (\bibinfo {year}
  {1980})}\BibitemShut {NoStop}%
\bibitem [{\citenamefont {Hatafuku}\ \emph {et~al.}(1983)\citenamefont
  {Hatafuku}, \citenamefont {Takahashi}, \citenamefont {Sasaki},\ and\
  \citenamefont {Ichinohe}}]{Hatafuku1983}%
  \BibitemOpen
  \bibfield  {author} {\bibinfo {author} {\bibfnamefont {H.}~\bibnamefont
  {Hatafuku}}, \bibinfo {author} {\bibfnamefont {S.}~\bibnamefont {Takahashi}},
  \bibinfo {author} {\bibfnamefont {T.}~\bibnamefont {Sasaki}}, \ and\ \bibinfo
  {author} {\bibfnamefont {H.}~\bibnamefont {Ichinohe}},\ }\href {\doibase
  https://doi.org/10.1016/0304-8853(83)90709-6} {\bibfield  {journal} {\bibinfo
   {journal} {J. Magn. Magn. Mater.}\ }\textbf {\bibinfo {volume} {31-34}},\
  \bibinfo {pages} {847} (\bibinfo {year} {1983})}\BibitemShut {NoStop}%
\bibitem [{\citenamefont {James}\ \emph {et~al.}(2000)\citenamefont {James},
  \citenamefont {Eriksson}, \citenamefont {Hjortstam}, \citenamefont
  {Johansson},\ and\ \citenamefont {Nordstr{\"o}m}}]{James2000}%
  \BibitemOpen
  \bibfield  {author} {\bibinfo {author} {\bibfnamefont {P.}~\bibnamefont
  {James}}, \bibinfo {author} {\bibfnamefont {O.}~\bibnamefont {Eriksson}},
  \bibinfo {author} {\bibfnamefont {O.}~\bibnamefont {Hjortstam}}, \bibinfo
  {author} {\bibfnamefont {B.}~\bibnamefont {Johansson}}, \ and\ \bibinfo
  {author} {\bibfnamefont {L.}~\bibnamefont {Nordstr{\"o}m}},\ }\href@noop {}
  {\bibfield  {journal} {\bibinfo  {journal} {Appl. Phys. Lett.}\ }\textbf
  {\bibinfo {volume} {76}},\ \bibinfo {pages} {915} (\bibinfo {year}
  {2000})}\BibitemShut {NoStop}%
\bibitem [{\citenamefont {Yin}\ \emph {et~al.}(2006)\citenamefont {Yin},
  \citenamefont {Wei}, \citenamefont {Lei}, \citenamefont {Zhou}, \citenamefont
  {Tian}, \citenamefont {Dong}, \citenamefont {Jin}, \citenamefont {Guo},
  \citenamefont {Jia},\ and\ \citenamefont {Wu}}]{Yin2006}%
  \BibitemOpen
  \bibfield  {author} {\bibinfo {author} {\bibfnamefont {L.~F.}\ \bibnamefont
  {Yin}}, \bibinfo {author} {\bibfnamefont {D.~H.}\ \bibnamefont {Wei}},
  \bibinfo {author} {\bibfnamefont {N.}~\bibnamefont {Lei}}, \bibinfo {author}
  {\bibfnamefont {L.~H.}\ \bibnamefont {Zhou}}, \bibinfo {author}
  {\bibfnamefont {C.~S.}\ \bibnamefont {Tian}}, \bibinfo {author}
  {\bibfnamefont {G.~S.}\ \bibnamefont {Dong}}, \bibinfo {author}
  {\bibfnamefont {X.~F.}\ \bibnamefont {Jin}}, \bibinfo {author} {\bibfnamefont
  {L.~P.}\ \bibnamefont {Guo}}, \bibinfo {author} {\bibfnamefont {Q.~J.}\
  \bibnamefont {Jia}}, \ and\ \bibinfo {author} {\bibfnamefont {R.~Q.}\
  \bibnamefont {Wu}},\ }\href@noop {} {\bibfield  {journal} {\bibinfo
  {journal} {Phys. Rev. Lett.}\ }\textbf {\bibinfo {volume} {97}},\ \bibinfo
  {pages} {067203} (\bibinfo {year} {2006})}\BibitemShut {NoStop}%
\bibitem [{\citenamefont {Ustinovshikov}\ and\ \citenamefont
  {Shabanova}(2013)}]{Ustinovshikov2013}%
  \BibitemOpen
  \bibfield  {author} {\bibinfo {author} {\bibfnamefont {Y.}~\bibnamefont
  {Ustinovshikov}}\ and\ \bibinfo {author} {\bibfnamefont {I.}~\bibnamefont
  {Shabanova}},\ }\href {\doibase
  https://doi.org/10.1016/j.jallcom.2013.06.039} {\bibfield  {journal}
  {\bibinfo  {journal} {J. Alloys Compd.}\ }\textbf {\bibinfo {volume} {578}},\
  \bibinfo {pages} {292} (\bibinfo {year} {2013})}\BibitemShut {NoStop}%
\bibitem [{\citenamefont {Waeckerl{\'e}}\ \emph {et~al.}(2020)\citenamefont
  {Waeckerl{\'e}}, \citenamefont {Demier}, \citenamefont {Godard},\ and\
  \citenamefont {Fraisse}}]{Waeckerle2020}%
  \BibitemOpen
  \bibfield  {author} {\bibinfo {author} {\bibfnamefont {T.}~\bibnamefont
  {Waeckerl{\'e}}}, \bibinfo {author} {\bibfnamefont {A.}~\bibnamefont
  {Demier}}, \bibinfo {author} {\bibfnamefont {F.}~\bibnamefont {Godard}}, \
  and\ \bibinfo {author} {\bibfnamefont {H.}~\bibnamefont {Fraisse}},\ }\href
  {\doibase https://doi.org/10.1016/j.jmmm.2020.166635} {\bibfield  {journal}
  {\bibinfo  {journal} {J. Magn. Magn. Mater.}\ }\textbf {\bibinfo {volume}
  {505}},\ \bibinfo {pages} {166635} (\bibinfo {year} {2020})}\BibitemShut
  {NoStop}%
\bibitem [{\citenamefont {Faulkner}(1982)}]{Faulkner1982}%
  \BibitemOpen
  \bibfield  {author} {\bibinfo {author} {\bibfnamefont {J.~S.}\ \bibnamefont
  {Faulkner}},\ }\href {\doibase https://doi.org/10.1016/0079-6425(82)90005-6}
  {\bibfield  {journal} {\bibinfo  {journal} {Prog. Mater. Sci.}\ }\textbf
  {\bibinfo {volume} {27}},\ \bibinfo {pages} {1} (\bibinfo {year}
  {1982})}\BibitemShut {NoStop}%
\bibitem [{\citenamefont {Feder}\ \emph {et~al.}(1983)\citenamefont {Feder},
  \citenamefont {Rosicky},\ and\ \citenamefont {Ackermann}}]{Feder1983}%
  \BibitemOpen
  \bibfield  {author} {\bibinfo {author} {\bibfnamefont {R.}~\bibnamefont
  {Feder}}, \bibinfo {author} {\bibfnamefont {F.}~\bibnamefont {Rosicky}}, \
  and\ \bibinfo {author} {\bibfnamefont {B.}~\bibnamefont {Ackermann}},\
  }\href@noop {} {\bibfield  {journal} {\bibinfo  {journal} {Z. Physik B --
  Condensed Matter}\ }\textbf {\bibinfo {volume} {52}},\ \bibinfo {pages} {31}
  (\bibinfo {year} {1983})}\BibitemShut {NoStop}%
\bibitem [{\citenamefont {Strange}\ \emph {et~al.}(1989)\citenamefont
  {Strange}, \citenamefont {Ebert}, \citenamefont {Staunton},\ and\
  \citenamefont {Gyorffy}}]{Strange1989}%
  \BibitemOpen
  \bibfield  {author} {\bibinfo {author} {\bibfnamefont {P.}~\bibnamefont
  {Strange}}, \bibinfo {author} {\bibfnamefont {H.}~\bibnamefont {Ebert}},
  \bibinfo {author} {\bibfnamefont {J.~B.}\ \bibnamefont {Staunton}}, \ and\
  \bibinfo {author} {\bibfnamefont {B.~L.}\ \bibnamefont {Gyorffy}},\
  }\href@noop {} {\bibfield  {journal} {\bibinfo  {journal} {J. Phys. Condens.
  Matter}\ }\textbf {\bibinfo {volume} {1}},\ \bibinfo {pages} {2959} (\bibinfo
  {year} {1989})}\BibitemShut {NoStop}%
\bibitem [{\citenamefont {Akai}(1989)}]{Akai1989}%
  \BibitemOpen
  \bibfield  {author} {\bibinfo {author} {\bibfnamefont {H.}~\bibnamefont
  {Akai}},\ }\href@noop {} {\bibfield  {journal} {\bibinfo  {journal} {J. Phys.
  Condens. Matter}\ }\textbf {\bibinfo {volume} {1}},\ \bibinfo {pages} {8045}
  (\bibinfo {year} {1989})}\BibitemShut {NoStop}%
\bibitem [{\citenamefont {Dederichs}\ \emph {et~al.}(1991)\citenamefont
  {Dederichs}, \citenamefont {Zeller}, \citenamefont {Akai},\ and\
  \citenamefont {Ebert}}]{Dederichs1991}%
  \BibitemOpen
  \bibfield  {author} {\bibinfo {author} {\bibfnamefont {P.~H.}\ \bibnamefont
  {Dederichs}}, \bibinfo {author} {\bibfnamefont {R.}~\bibnamefont {Zeller}},
  \bibinfo {author} {\bibfnamefont {H.}~\bibnamefont {Akai}}, \ and\ \bibinfo
  {author} {\bibfnamefont {H.}~\bibnamefont {Ebert}},\ }\href {\doibase
  https://doi.org/10.1016/0304-8853(91)90823-S} {\bibfield  {journal} {\bibinfo
   {journal} {J. Magn. Magn. Mater.}\ }\textbf {\bibinfo {volume} {100}},\
  \bibinfo {pages} {241} (\bibinfo {year} {1991})}\BibitemShut {NoStop}%
\bibitem [{\citenamefont {Ebert}\ \emph {et~al.}(1992)\citenamefont {Ebert},
  \citenamefont {Drittler},\ and\ \citenamefont {Akai}}]{Ebert1992}%
  \BibitemOpen
  \bibfield  {author} {\bibinfo {author} {\bibfnamefont {H.}~\bibnamefont
  {Ebert}}, \bibinfo {author} {\bibfnamefont {B.}~\bibnamefont {Drittler}}, \
  and\ \bibinfo {author} {\bibfnamefont {H.}~\bibnamefont {Akai}},\ }\href
  {\doibase https://doi.org/10.1016/0304-8853(92)91007-G} {\bibfield  {journal}
  {\bibinfo  {journal} {J. Magn. Magn. Mater.}\ }\textbf {\bibinfo {volume}
  {104-107}},\ \bibinfo {pages} {733} (\bibinfo {year} {1992})}\BibitemShut
  {NoStop}%
\bibitem [{\citenamefont {Rowlands}\ \emph {et~al.}(2003)\citenamefont
  {Rowlands}, \citenamefont {Staunton},\ and\ \citenamefont
  {Gy\"orffy}}]{Rowlands2003}%
  \BibitemOpen
  \bibfield  {author} {\bibinfo {author} {\bibfnamefont {D.~A.}\ \bibnamefont
  {Rowlands}}, \bibinfo {author} {\bibfnamefont {J.~B.}\ \bibnamefont
  {Staunton}}, \ and\ \bibinfo {author} {\bibfnamefont {B.~L.}\ \bibnamefont
  {Gy\"orffy}},\ }\href {\doibase 10.1103/PhysRevB.67.115109} {\bibfield
  {journal} {\bibinfo  {journal} {Phys. Rev. B}\ }\textbf {\bibinfo {volume}
  {67}},\ \bibinfo {pages} {115109} (\bibinfo {year} {2003})}\BibitemShut
  {NoStop}%
\bibitem [{\citenamefont {Rowlands}\ \emph {et~al.}(2005)\citenamefont
  {Rowlands}, \citenamefont {Staunton}, \citenamefont {Gy\"orffy},
  \citenamefont {Bruno},\ and\ \citenamefont {Ginatempo}}]{Rowlands2005}%
  \BibitemOpen
  \bibfield  {author} {\bibinfo {author} {\bibfnamefont {D.~A.}\ \bibnamefont
  {Rowlands}}, \bibinfo {author} {\bibfnamefont {J.~B.}\ \bibnamefont
  {Staunton}}, \bibinfo {author} {\bibfnamefont {B.~L.}\ \bibnamefont
  {Gy\"orffy}}, \bibinfo {author} {\bibfnamefont {E.}~\bibnamefont {Bruno}}, \
  and\ \bibinfo {author} {\bibfnamefont {B.}~\bibnamefont {Ginatempo}},\ }\href
  {\doibase 10.1103/PhysRevB.72.045101} {\bibfield  {journal} {\bibinfo
  {journal} {Phys. Rev. B}\ }\textbf {\bibinfo {volume} {72}},\ \bibinfo
  {pages} {045101} (\bibinfo {year} {2005})}\BibitemShut {NoStop}%
\bibitem [{\citenamefont {Rowlands}(2009)}]{Row09}%
  \BibitemOpen
  \bibfield  {author} {\bibinfo {author} {\bibfnamefont {D.~A.}\ \bibnamefont
  {Rowlands}},\ }\href@noop {} {\bibfield  {journal} {\bibinfo  {journal}
  {Reports on Progress in Physics}\ }\textbf {\bibinfo {volume} {72}},\
  \bibinfo {pages} {086501} (\bibinfo {year} {2009})}\BibitemShut {NoStop}%
\bibitem [{\citenamefont {Ebert}\ \emph {et~al.}(2003)\citenamefont {Ebert},
  \citenamefont {Perlov},\ and\ \citenamefont {Mankovsky}}]{Ebert2003}%
  \BibitemOpen
  \bibfield  {author} {\bibinfo {author} {\bibfnamefont {H.}~\bibnamefont
  {Ebert}}, \bibinfo {author} {\bibfnamefont {A.}~\bibnamefont {Perlov}}, \
  and\ \bibinfo {author} {\bibfnamefont {S.}~\bibnamefont {Mankovsky}},\ }\href
  {\doibase https://doi.org/10.1016/S0038-1098(03)00455-1} {\bibfield
  {journal} {\bibinfo  {journal} {Solid State Commun.}\ }\textbf {\bibinfo
  {volume} {127}},\ \bibinfo {pages} {443} (\bibinfo {year}
  {2003})}\BibitemShut {NoStop}%
\bibitem [{\citenamefont {Min\'ar}\ \emph {et~al.}(2005)\citenamefont
  {Min\'ar}, \citenamefont {Chioncel}, \citenamefont {Perlov}, \citenamefont
  {Ebert}, \citenamefont {Katsnelson},\ and\ \citenamefont
  {Lichtenstein}}]{Minar2005}%
  \BibitemOpen
  \bibfield  {author} {\bibinfo {author} {\bibfnamefont {J.}~\bibnamefont
  {Min\'ar}}, \bibinfo {author} {\bibfnamefont {L.}~\bibnamefont {Chioncel}},
  \bibinfo {author} {\bibfnamefont {A.}~\bibnamefont {Perlov}}, \bibinfo
  {author} {\bibfnamefont {H.}~\bibnamefont {Ebert}}, \bibinfo {author}
  {\bibfnamefont {M.~I.}\ \bibnamefont {Katsnelson}}, \ and\ \bibinfo {author}
  {\bibfnamefont {A.~I.}\ \bibnamefont {Lichtenstein}},\ }\href {\doibase
  10.1103/PhysRevB.72.045125} {\bibfield  {journal} {\bibinfo  {journal} {Phys.
  Rev. B}\ }\textbf {\bibinfo {volume} {72}},\ \bibinfo {pages} {045125}
  (\bibinfo {year} {2005})}\BibitemShut {NoStop}%
\bibitem [{\citenamefont {\ifmmode~\check{S}\else \v{S}\fi{}ipr}\ \emph
  {et~al.}(2008)\citenamefont {\ifmmode~\check{S}\else \v{S}\fi{}ipr},
  \citenamefont {Min\'ar}, \citenamefont {Mankovsky},\ and\ \citenamefont
  {Ebert}}]{Sipr2008}%
  \BibitemOpen
  \bibfield  {author} {\bibinfo {author} {\bibfnamefont {O.}~\bibnamefont
  {\ifmmode~\check{S}\else \v{S}\fi{}ipr}}, \bibinfo {author} {\bibfnamefont
  {J.}~\bibnamefont {Min\'ar}}, \bibinfo {author} {\bibfnamefont
  {S.}~\bibnamefont {Mankovsky}}, \ and\ \bibinfo {author} {\bibfnamefont
  {H.}~\bibnamefont {Ebert}},\ }\href {\doibase 10.1103/PhysRevB.78.144403}
  {\bibfield  {journal} {\bibinfo  {journal} {Phys. Rev. B}\ }\textbf {\bibinfo
  {volume} {78}},\ \bibinfo {pages} {144403} (\bibinfo {year}
  {2008})}\BibitemShut {NoStop}%
\bibitem [{\citenamefont {Braun}\ \emph {et~al.}(2010)\citenamefont {Braun},
  \citenamefont {Min\'ar}, \citenamefont {Matthes}, \citenamefont {Schneider},\
  and\ \citenamefont {Ebert}}]{Braun2010}%
  \BibitemOpen
  \bibfield  {author} {\bibinfo {author} {\bibfnamefont {J.}~\bibnamefont
  {Braun}}, \bibinfo {author} {\bibfnamefont {J.}~\bibnamefont {Min\'ar}},
  \bibinfo {author} {\bibfnamefont {F.}~\bibnamefont {Matthes}}, \bibinfo
  {author} {\bibfnamefont {C.~M.}\ \bibnamefont {Schneider}}, \ and\ \bibinfo
  {author} {\bibfnamefont {H.}~\bibnamefont {Ebert}},\ }\href {\doibase
  10.1103/PhysRevB.82.024411} {\bibfield  {journal} {\bibinfo  {journal} {Phys.
  Rev. B}\ }\textbf {\bibinfo {volume} {82}},\ \bibinfo {pages} {024411}
  (\bibinfo {year} {2010})}\BibitemShut {NoStop}%
\bibitem [{\citenamefont {Min{\'{a}}r}\ \emph {et~al.}(2014)\citenamefont
  {Min{\'{a}}r}, \citenamefont {Mankovsky}, \citenamefont {{\v{S}}ipr},
  \citenamefont {Benea},\ and\ \citenamefont {Ebert}}]{Minar2014}%
  \BibitemOpen
  \bibfield  {author} {\bibinfo {author} {\bibfnamefont {J.}~\bibnamefont
  {Min{\'{a}}r}}, \bibinfo {author} {\bibfnamefont {S.}~\bibnamefont
  {Mankovsky}}, \bibinfo {author} {\bibfnamefont {O.}~\bibnamefont
  {{\v{S}}ipr}}, \bibinfo {author} {\bibfnamefont {D.}~\bibnamefont {Benea}}, \
  and\ \bibinfo {author} {\bibfnamefont {H.}~\bibnamefont {Ebert}},\ }\href
  {\doibase 10.1088/0953-8984/26/27/274206} {\bibfield  {journal} {\bibinfo
  {journal} {J. Phys. Condens. Matter}\ }\textbf {\bibinfo {volume} {26}},\
  \bibinfo {pages} {274206} (\bibinfo {year} {2014})}\BibitemShut {NoStop}%
\bibitem [{\citenamefont {Poteryaev}\ \emph {et~al.}(2016)\citenamefont
  {Poteryaev}, \citenamefont {Skorikov}, \citenamefont {Anisimov},\ and\
  \citenamefont {Korotin}}]{Poteryaev2016}%
  \BibitemOpen
  \bibfield  {author} {\bibinfo {author} {\bibfnamefont {A.~I.}\ \bibnamefont
  {Poteryaev}}, \bibinfo {author} {\bibfnamefont {N.~A.}\ \bibnamefont
  {Skorikov}}, \bibinfo {author} {\bibfnamefont {V.~I.}\ \bibnamefont
  {Anisimov}}, \ and\ \bibinfo {author} {\bibfnamefont {M.~A.}\ \bibnamefont
  {Korotin}},\ }\href {\doibase 10.1103/PhysRevB.93.205135} {\bibfield
  {journal} {\bibinfo  {journal} {Phys. Rev. B}\ }\textbf {\bibinfo {volume}
  {93}},\ \bibinfo {pages} {205135} (\bibinfo {year} {2016})}\BibitemShut
  {NoStop}%
\bibitem [{\citenamefont {Faulkner}\ \emph {et~al.}(2001)\citenamefont
  {Faulkner}, \citenamefont {Ujfalussy}, \citenamefont {Moghadam},
  \citenamefont {Stocks},\ and\ \citenamefont {Wang}}]{FUM01}%
  \BibitemOpen
  \bibfield  {author} {\bibinfo {author} {\bibfnamefont {J.~S.}\ \bibnamefont
  {Faulkner}}, \bibinfo {author} {\bibfnamefont {B.}~\bibnamefont {Ujfalussy}},
  \bibinfo {author} {\bibfnamefont {N.}~\bibnamefont {Moghadam}}, \bibinfo
  {author} {\bibfnamefont {G.~M.}\ \bibnamefont {Stocks}}, \ and\ \bibinfo
  {author} {\bibfnamefont {Y.}~\bibnamefont {Wang}},\ }\href@noop {} {\bibfield
   {journal} {\bibinfo  {journal} {J. Phys. Condens. Matter}\ }\textbf
  {\bibinfo {volume} {13}},\ \bibinfo {pages} {8573} (\bibinfo {year}
  {2001})}\BibitemShut {NoStop}%
\bibitem [{\citenamefont {Popescu}\ and\ \citenamefont
  {Zunger}(2010)}]{Popescu2010}%
  \BibitemOpen
  \bibfield  {author} {\bibinfo {author} {\bibfnamefont {V.}~\bibnamefont
  {Popescu}}\ and\ \bibinfo {author} {\bibfnamefont {A.}~\bibnamefont
  {Zunger}},\ }\href {\doibase 10.1103/PhysRevLett.104.236403} {\bibfield
  {journal} {\bibinfo  {journal} {Phys. Rev. Lett.}\ }\textbf {\bibinfo
  {volume} {104}},\ \bibinfo {pages} {236403} (\bibinfo {year}
  {2010})}\BibitemShut {NoStop}%
\bibitem [{\citenamefont {Mano}(1977)}]{Mano1977}%
  \BibitemOpen
  \bibfield  {author} {\bibinfo {author} {\bibfnamefont {H.}~\bibnamefont
  {Mano}},\ }\href {\doibase 10.1143/PTP.57.1848} {\bibfield  {journal}
  {\bibinfo  {journal} {Prog. Theor. Phys.}\ }\textbf {\bibinfo {volume}
  {57}},\ \bibinfo {pages} {1848} (\bibinfo {year} {1977})}\BibitemShut
  {NoStop}%
\bibitem [{\citenamefont {Liechtenstein}\ \emph {et~al.}(1987)\citenamefont
  {Liechtenstein}, \citenamefont {Katsnelson}, \citenamefont {Antropov},\ and\
  \citenamefont {Gubanov}}]{Liechtenstein1987}%
  \BibitemOpen
  \bibfield  {author} {\bibinfo {author} {\bibfnamefont {A.~I.}\ \bibnamefont
  {Liechtenstein}}, \bibinfo {author} {\bibfnamefont {M.~I.}\ \bibnamefont
  {Katsnelson}}, \bibinfo {author} {\bibfnamefont {V.~P.}\ \bibnamefont
  {Antropov}}, \ and\ \bibinfo {author} {\bibfnamefont {V.~A.}\ \bibnamefont
  {Gubanov}},\ }\href {\doibase https://doi.org/10.1016/0304-8853(87)90721-9}
  {\bibfield  {journal} {\bibinfo  {journal} {J. Magn. Magn. Mater.}\ }\textbf
  {\bibinfo {volume} {67}},\ \bibinfo {pages} {65} (\bibinfo {year}
  {1987})}\BibitemShut {NoStop}%
\bibitem [{\citenamefont {Halilov}\ \emph {et~al.}(1998)\citenamefont
  {Halilov}, \citenamefont {Eschrig}, \citenamefont {Perlov},\ and\
  \citenamefont {Oppeneer}}]{Halilov1998}%
  \BibitemOpen
  \bibfield  {author} {\bibinfo {author} {\bibfnamefont {S.~V.}\ \bibnamefont
  {Halilov}}, \bibinfo {author} {\bibfnamefont {H.}~\bibnamefont {Eschrig}},
  \bibinfo {author} {\bibfnamefont {A.~Y.}\ \bibnamefont {Perlov}}, \ and\
  \bibinfo {author} {\bibfnamefont {P.~M.}\ \bibnamefont {Oppeneer}},\ }\href
  {\doibase 10.1103/PhysRevB.58.293} {\bibfield  {journal} {\bibinfo  {journal}
  {Phys. Rev. B}\ }\textbf {\bibinfo {volume} {58}},\ \bibinfo {pages} {293}
  (\bibinfo {year} {1998})}\BibitemShut {NoStop}%
\bibitem [{\citenamefont {Turek}\ \emph {et~al.}(2006)\citenamefont {Turek},
  \citenamefont {Kudrnovsk{\'y}}, \citenamefont {Drchal},\ and\ \citenamefont
  {Bruno}}]{Turek2006}%
  \BibitemOpen
  \bibfield  {author} {\bibinfo {author} {\bibfnamefont {I.}~\bibnamefont
  {Turek}}, \bibinfo {author} {\bibfnamefont {J.}~\bibnamefont
  {Kudrnovsk{\'y}}}, \bibinfo {author} {\bibfnamefont {V.}~\bibnamefont
  {Drchal}}, \ and\ \bibinfo {author} {\bibfnamefont {P.}~\bibnamefont
  {Bruno}},\ }\href@noop {} {\bibfield  {journal} {\bibinfo  {journal} {Philos.
  Mag.}\ }\textbf {\bibinfo {volume} {86}},\ \bibinfo {pages} {1713} (\bibinfo
  {year} {2006})}\BibitemShut {NoStop}%
\bibitem [{\citenamefont {Takahashi}\ \emph {et~al.}(2007)\citenamefont
  {Takahashi}, \citenamefont {Ogura},\ and\ \citenamefont
  {Akai}}]{Takahashi2007}%
  \BibitemOpen
  \bibfield  {author} {\bibinfo {author} {\bibfnamefont {C.}~\bibnamefont
  {Takahashi}}, \bibinfo {author} {\bibfnamefont {M.}~\bibnamefont {Ogura}}, \
  and\ \bibinfo {author} {\bibfnamefont {H.}~\bibnamefont {Akai}},\ }\href@noop
  {} {\bibfield  {journal} {\bibinfo  {journal} {J. Phys. Condens. Matter}\
  }\textbf {\bibinfo {volume} {19}},\ \bibinfo {pages} {365233} (\bibinfo
  {year} {2007})}\BibitemShut {NoStop}%
\bibitem [{\citenamefont {Evans}(2018)}]{Evans2020}%
  \BibitemOpen
  \bibfield  {author} {\bibinfo {author} {\bibfnamefont {R.~F.~L.}\
  \bibnamefont {Evans}},\ }\enquote {\bibinfo {title} {Atomistic spin
  dynamics},}\ in\ \href {\doibase 10.1007/978-3-319-50257-1_147-1} {\emph
  {\bibinfo {booktitle} {Handbook of Materials Modeling: Applications: Current
  and Emerging Materials}}},\ \bibinfo {editor} {edited by\ \bibinfo {editor}
  {\bibfnamefont {W.}~\bibnamefont {Andreoni}}\ and\ \bibinfo {editor}
  {\bibfnamefont {S.}~\bibnamefont {Yip}}}\ (\bibinfo  {publisher} {Springer},\
  \bibinfo {address} {Cham},\ \bibinfo {year} {2018})\ pp.\ \bibinfo {pages}
  {1--23}\BibitemShut {NoStop}%
\bibitem [{\citenamefont {Pindor}\ \emph {et~al.}(1983)\citenamefont {Pindor},
  \citenamefont {Staunton}, \citenamefont {Stocks},\ and\ \citenamefont
  {Winter}}]{Pindor1983}%
  \BibitemOpen
  \bibfield  {author} {\bibinfo {author} {\bibfnamefont {A.~J.}\ \bibnamefont
  {Pindor}}, \bibinfo {author} {\bibfnamefont {J.}~\bibnamefont {Staunton}},
  \bibinfo {author} {\bibfnamefont {G.~M.}\ \bibnamefont {Stocks}}, \ and\
  \bibinfo {author} {\bibfnamefont {H.}~\bibnamefont {Winter}},\ }\href
  {\doibase 10.1088/0305-4608/13/5/012} {\ \textbf {\bibinfo {volume} {13}},\
  \bibinfo {pages} {979} (\bibinfo {year} {1983})}\BibitemShut {NoStop}%
\bibitem [{\citenamefont {Gyorffy}\ \emph {et~al.}(1985)\citenamefont
  {Gyorffy}, \citenamefont {Pindor}, \citenamefont {Staunton}, \citenamefont
  {Stocks},\ and\ \citenamefont {Winter}}]{Gyorffy1985}%
  \BibitemOpen
  \bibfield  {author} {\bibinfo {author} {\bibfnamefont {B.~L.}\ \bibnamefont
  {Gyorffy}}, \bibinfo {author} {\bibfnamefont {A.~J.}\ \bibnamefont {Pindor}},
  \bibinfo {author} {\bibfnamefont {J.}~\bibnamefont {Staunton}}, \bibinfo
  {author} {\bibfnamefont {G.~M.}\ \bibnamefont {Stocks}}, \ and\ \bibinfo
  {author} {\bibfnamefont {H.}~\bibnamefont {Winter}},\ }\href {\doibase
  10.1088/0305-4608/15/6/018} {\ \textbf {\bibinfo {volume} {15}},\ \bibinfo
  {pages} {1337} (\bibinfo {year} {1985})}\BibitemShut {NoStop}%
\bibitem [{\citenamefont {Staunton}\ and\ \citenamefont
  {Gyorffy}(1992)}]{Staunton1992}%
  \BibitemOpen
  \bibfield  {author} {\bibinfo {author} {\bibfnamefont {J.~B.}\ \bibnamefont
  {Staunton}}\ and\ \bibinfo {author} {\bibfnamefont {B.~L.}\ \bibnamefont
  {Gyorffy}},\ }\href {\doibase 10.1103/PhysRevLett.69.371} {\bibfield
  {journal} {\bibinfo  {journal} {Phys. Rev. Lett.}\ }\textbf {\bibinfo
  {volume} {69}},\ \bibinfo {pages} {371} (\bibinfo {year} {1992})}\BibitemShut
  {NoStop}%
\bibitem [{\citenamefont {Staunton}\ \emph {et~al.}(2004)\citenamefont
  {Staunton}, \citenamefont {Ostanin}, \citenamefont {Razee}, \citenamefont
  {Gyorffy}, \citenamefont {Szunyogh}, \citenamefont {Ginatempo},\ and\
  \citenamefont {Bruno}}]{Staunton2004}%
  \BibitemOpen
  \bibfield  {author} {\bibinfo {author} {\bibfnamefont {J.~B.}\ \bibnamefont
  {Staunton}}, \bibinfo {author} {\bibfnamefont {S.}~\bibnamefont {Ostanin}},
  \bibinfo {author} {\bibfnamefont {S.~S.~A.}\ \bibnamefont {Razee}}, \bibinfo
  {author} {\bibfnamefont {B.~L.}\ \bibnamefont {Gyorffy}}, \bibinfo {author}
  {\bibfnamefont {L.}~\bibnamefont {Szunyogh}}, \bibinfo {author}
  {\bibfnamefont {B.}~\bibnamefont {Ginatempo}}, \ and\ \bibinfo {author}
  {\bibfnamefont {E.}~\bibnamefont {Bruno}},\ }\href {\doibase
  10.1103/PhysRevLett.93.257204} {\bibfield  {journal} {\bibinfo  {journal}
  {Phys. Rev. Lett.}\ }\textbf {\bibinfo {volume} {93}},\ \bibinfo {pages}
  {257204} (\bibinfo {year} {2004})}\BibitemShut {NoStop}%
\bibitem [{\citenamefont {Kudrnovsk\'y}\ \emph {et~al.}(2012)\citenamefont
  {Kudrnovsk\'y}, \citenamefont {Drchal}, \citenamefont {Turek}, \citenamefont
  {Khmelevskyi}, \citenamefont {Glasbrenner},\ and\ \citenamefont
  {Belashchenko}}]{Kudrnovsky2012}%
  \BibitemOpen
  \bibfield  {author} {\bibinfo {author} {\bibfnamefont {J.}~\bibnamefont
  {Kudrnovsk\'y}}, \bibinfo {author} {\bibfnamefont {V.}~\bibnamefont
  {Drchal}}, \bibinfo {author} {\bibfnamefont {I.}~\bibnamefont {Turek}},
  \bibinfo {author} {\bibfnamefont {S.}~\bibnamefont {Khmelevskyi}}, \bibinfo
  {author} {\bibfnamefont {J.~K.}\ \bibnamefont {Glasbrenner}}, \ and\ \bibinfo
  {author} {\bibfnamefont {K.~D.}\ \bibnamefont {Belashchenko}},\ }\href
  {\doibase 10.1103/PhysRevB.86.144423} {\bibfield  {journal} {\bibinfo
  {journal} {Phys. Rev. B}\ }\textbf {\bibinfo {volume} {86}},\ \bibinfo
  {pages} {144423} (\bibinfo {year} {2012})}\BibitemShut {NoStop}%
\bibitem [{\citenamefont {Hasegawa}(1981)}]{Hasegawa1981}%
  \BibitemOpen
  \bibfield  {author} {\bibinfo {author} {\bibfnamefont {H.}~\bibnamefont
  {Hasegawa}},\ }in\ \href@noop {} {\emph {\bibinfo {booktitle} {Electron
  correlation and magnetism in narrow-band systems}}}\ (\bibinfo  {publisher}
  {Springer},\ \bibinfo {address} {Berlin},\ \bibinfo {year} {1981})\ pp.\
  \bibinfo {pages} {38--50}\BibitemShut {NoStop}%
\bibitem [{\citenamefont {Uhl}\ and\ \citenamefont {K\"ubler}(1996)}]{Uhl1996}%
  \BibitemOpen
  \bibfield  {author} {\bibinfo {author} {\bibfnamefont {M.}~\bibnamefont
  {Uhl}}\ and\ \bibinfo {author} {\bibfnamefont {J.}~\bibnamefont {K\"ubler}},\
  }\href {\doibase 10.1103/PhysRevLett.77.334} {\bibfield  {journal} {\bibinfo
  {journal} {Phys. Rev. Lett.}\ }\textbf {\bibinfo {volume} {77}},\ \bibinfo
  {pages} {334} (\bibinfo {year} {1996})}\BibitemShut {NoStop}%
\bibitem [{\citenamefont {Rosengaard}\ and\ \citenamefont
  {Johansson}(1997)}]{Rosengaard1997}%
  \BibitemOpen
  \bibfield  {author} {\bibinfo {author} {\bibfnamefont {N.~M.}\ \bibnamefont
  {Rosengaard}}\ and\ \bibinfo {author} {\bibfnamefont {B.}~\bibnamefont
  {Johansson}},\ }\href {\doibase 10.1103/PhysRevB.55.14975} {\bibfield
  {journal} {\bibinfo  {journal} {Phys. Rev. B}\ }\textbf {\bibinfo {volume}
  {55}},\ \bibinfo {pages} {14975} (\bibinfo {year} {1997})}\BibitemShut
  {NoStop}%
\bibitem [{\citenamefont {Kakehashi}(2002)}]{Kakehashi2002}%
  \BibitemOpen
  \bibfield  {author} {\bibinfo {author} {\bibfnamefont {Y.}~\bibnamefont
  {Kakehashi}},\ }\href {\doibase 10.1103/PhysRevB.65.184420} {\bibfield
  {journal} {\bibinfo  {journal} {Phys. Rev. B}\ }\textbf {\bibinfo {volume}
  {65}},\ \bibinfo {pages} {184420} (\bibinfo {year} {2002})}\BibitemShut
  {NoStop}%
\bibitem [{\citenamefont {Shallcross}\ \emph {et~al.}(2005)\citenamefont
  {Shallcross}, \citenamefont {Kissavos}, \citenamefont {Meded},\ and\
  \citenamefont {Ruban}}]{Shallcross2005}%
  \BibitemOpen
  \bibfield  {author} {\bibinfo {author} {\bibfnamefont {S.}~\bibnamefont
  {Shallcross}}, \bibinfo {author} {\bibfnamefont {A.~E.}\ \bibnamefont
  {Kissavos}}, \bibinfo {author} {\bibfnamefont {V.}~\bibnamefont {Meded}}, \
  and\ \bibinfo {author} {\bibfnamefont {A.~V.}\ \bibnamefont {Ruban}},\ }\href
  {\doibase 10.1103/PhysRevB.72.104437} {\bibfield  {journal} {\bibinfo
  {journal} {Phys. Rev. B}\ }\textbf {\bibinfo {volume} {72}},\ \bibinfo
  {pages} {104437} (\bibinfo {year} {2005})}\BibitemShut {NoStop}%
\bibitem [{\citenamefont {Ruban}\ \emph {et~al.}(2007)\citenamefont {Ruban},
  \citenamefont {Khmelevskyi}, \citenamefont {Mohn},\ and\ \citenamefont
  {Johansson}}]{Ruban2007}%
  \BibitemOpen
  \bibfield  {author} {\bibinfo {author} {\bibfnamefont {A.~V.}\ \bibnamefont
  {Ruban}}, \bibinfo {author} {\bibfnamefont {S.}~\bibnamefont {Khmelevskyi}},
  \bibinfo {author} {\bibfnamefont {P.}~\bibnamefont {Mohn}}, \ and\ \bibinfo
  {author} {\bibfnamefont {B.}~\bibnamefont {Johansson}},\ }\href {\doibase
  10.1103/PhysRevB.75.054402} {\bibfield  {journal} {\bibinfo  {journal} {Phys.
  Rev. B}\ }\textbf {\bibinfo {volume} {75}},\ \bibinfo {pages} {054402}
  (\bibinfo {year} {2007})}\BibitemShut {NoStop}%
\bibitem [{\citenamefont {Ma}\ and\ \citenamefont {Dudarev}(2012)}]{MD12}%
  \BibitemOpen
  \bibfield  {author} {\bibinfo {author} {\bibfnamefont {P.-W.}\ \bibnamefont
  {Ma}}\ and\ \bibinfo {author} {\bibfnamefont {S.~L.}\ \bibnamefont
  {Dudarev}},\ }\href {\doibase 10.1103/PhysRevB.86.054416} {\bibfield
  {journal} {\bibinfo  {journal} {Phys. Rev. B}\ }\textbf {\bibinfo {volume}
  {86}},\ \bibinfo {pages} {054416} (\bibinfo {year} {2012})}\BibitemShut
  {NoStop}%
\bibitem [{\citenamefont {Grebennikov}\ and\ \citenamefont
  {Radzivonchik}(2015)}]{Grebennikov2015}%
  \BibitemOpen
  \bibfield  {author} {\bibinfo {author} {\bibfnamefont {V.~I.}\ \bibnamefont
  {Grebennikov}}\ and\ \bibinfo {author} {\bibfnamefont {D.~I.}\ \bibnamefont
  {Radzivonchik}},\ }in\ \href@noop {} {\emph {\bibinfo {booktitle} {Solid
  State Phenom.}}},\ Vol.\ \bibinfo {volume} {233}\ (\bibinfo {organization}
  {Trans Tech Publ},\ \bibinfo {year} {2015})\ pp.\ \bibinfo {pages}
  {25--29}\BibitemShut {NoStop}%
\bibitem [{\citenamefont {Ruban}\ and\ \citenamefont
  {Dehghani}(2016)}]{Ruban2016}%
  \BibitemOpen
  \bibfield  {author} {\bibinfo {author} {\bibfnamefont {A.~V.}\ \bibnamefont
  {Ruban}}\ and\ \bibinfo {author} {\bibfnamefont {M.}~\bibnamefont
  {Dehghani}},\ }\href {\doibase 10.1103/PhysRevB.94.104111} {\bibfield
  {journal} {\bibinfo  {journal} {Phys. Rev. B}\ }\textbf {\bibinfo {volume}
  {94}},\ \bibinfo {pages} {104111} (\bibinfo {year} {2016})}\BibitemShut
  {NoStop}%
\bibitem [{\citenamefont {Ruban}(2017)}]{Ruban2017}%
  \BibitemOpen
  \bibfield  {author} {\bibinfo {author} {\bibfnamefont {A.~V.}\ \bibnamefont
  {Ruban}},\ }\href {\doibase 10.1103/PhysRevB.95.174432} {\bibfield  {journal}
  {\bibinfo  {journal} {Phys. Rev. B}\ }\textbf {\bibinfo {volume} {95}},\
  \bibinfo {pages} {174432} (\bibinfo {year} {2017})}\BibitemShut {NoStop}%
\bibitem [{\citenamefont {Pan}\ \emph {et~al.}(2017)\citenamefont {Pan},
  \citenamefont {Chico}, \citenamefont {Delin}, \citenamefont {Bergman},\ and\
  \citenamefont {Bergqvist}}]{Pan2017}%
  \BibitemOpen
  \bibfield  {author} {\bibinfo {author} {\bibfnamefont {F.}~\bibnamefont
  {Pan}}, \bibinfo {author} {\bibfnamefont {J.}~\bibnamefont {Chico}}, \bibinfo
  {author} {\bibfnamefont {A.}~\bibnamefont {Delin}}, \bibinfo {author}
  {\bibfnamefont {A.}~\bibnamefont {Bergman}}, \ and\ \bibinfo {author}
  {\bibfnamefont {L.}~\bibnamefont {Bergqvist}},\ }\href {\doibase
  10.1103/PhysRevB.95.184432} {\bibfield  {journal} {\bibinfo  {journal} {Phys.
  Rev. B}\ }\textbf {\bibinfo {volume} {95}},\ \bibinfo {pages} {184432}
  (\bibinfo {year} {2017})}\BibitemShut {NoStop}%
\bibitem [{\citenamefont {Melnikov}\ and\ \citenamefont {Reser}(2018)}]{MR18}%
  \BibitemOpen
  \bibfield  {author} {\bibinfo {author} {\bibfnamefont {N.~B.}\ \bibnamefont
  {Melnikov}}\ and\ \bibinfo {author} {\bibfnamefont {B.~I.}\ \bibnamefont
  {Reser}},\ }\href@noop {} {\emph {\bibinfo {title} {{Dynamic spin-fluctuation
  theory of metallic magnetism}}}}\ (\bibinfo  {publisher} {Springer},\
  \bibinfo {address} {Berlin},\ \bibinfo {year} {2018})\BibitemShut {NoStop}%
\bibitem [{\citenamefont {Ebert}\ \emph {et~al.}(2015)\citenamefont {Ebert},
  \citenamefont {Mankovsky}, \citenamefont {Chadova}, \citenamefont {Polesya},
  \citenamefont {Minar},\ and\ \citenamefont {Koedderitzsch}}]{Ebert2015}%
  \BibitemOpen
  \bibfield  {author} {\bibinfo {author} {\bibfnamefont {H.}~\bibnamefont
  {Ebert}}, \bibinfo {author} {\bibfnamefont {S.}~\bibnamefont {Mankovsky}},
  \bibinfo {author} {\bibfnamefont {K.}~\bibnamefont {Chadova}}, \bibinfo
  {author} {\bibfnamefont {S.}~\bibnamefont {Polesya}}, \bibinfo {author}
  {\bibfnamefont {J.}~\bibnamefont {Minar}}, \ and\ \bibinfo {author}
  {\bibfnamefont {D.}~\bibnamefont {Koedderitzsch}},\ }\href@noop {} {\bibfield
   {journal} {\bibinfo  {journal} {Phys. Rev. B}\ }\textbf {\bibinfo {volume}
  {91}},\ \bibinfo {pages} {165132} (\bibinfo {year} {2015})}\BibitemShut
  {NoStop}%
\bibitem [{\citenamefont {Ebert}\ \emph {et~al.}()\citenamefont {Ebert} \emph
  {et~al.}}]{SPRKKR}%
  \BibitemOpen
  \bibfield  {author} {\bibinfo {author} {\bibfnamefont {H.}~\bibnamefont
  {Ebert}} \emph {et~al.},\ }\href {http://ebert.cup.uni-muenchen.de/SPRKKR}
  {\enquote {\bibinfo {title} {{The Munich SPR-KKR package, version 7.7}},}\
  }\bibinfo {note} {Available online at
  \url{http://ebert.cup.uni-muenchen.de/SPRKKR}}\BibitemShut {NoStop}%
\bibitem [{\citenamefont {Ebert}\ \emph {et~al.}(2011)\citenamefont {Ebert},
  \citenamefont {K{\"o}edderitzsch},\ and\ \citenamefont
  {Min{\'a}r}}]{SPRKKR1}%
  \BibitemOpen
  \bibfield  {author} {\bibinfo {author} {\bibfnamefont {H.}~\bibnamefont
  {Ebert}}, \bibinfo {author} {\bibfnamefont {D.}~\bibnamefont
  {K{\"o}edderitzsch}}, \ and\ \bibinfo {author} {\bibfnamefont
  {J.}~\bibnamefont {Min{\'a}r}},\ }\href@noop {} {\bibfield  {journal}
  {\bibinfo  {journal} {Rep. Prog. Phys.}\ }\textbf {\bibinfo {volume} {74}},\
  \bibinfo {pages} {096501} (\bibinfo {year} {2011})}\BibitemShut {NoStop}%
\bibitem [{\citenamefont {Broyden}(1965)}]{Bro65}%
  \BibitemOpen
  \bibfield  {author} {\bibinfo {author} {\bibfnamefont {C.~G.}\ \bibnamefont
  {Broyden}},\ }\href@noop {} {\bibfield  {journal} {\bibinfo  {journal} {Math.
  Comp.}\ }\textbf {\bibinfo {volume} {19}},\ \bibinfo {pages} {577} (\bibinfo
  {year} {1965})}\BibitemShut {NoStop}%
\bibitem [{\citenamefont {Dennis}\ and\ \citenamefont {Schnabel}(1996)}]{DS96}%
  \BibitemOpen
  \bibfield  {author} {\bibinfo {author} {\bibfnamefont {J.~E.}\ \bibnamefont
  {Dennis}}\ and\ \bibinfo {author} {\bibfnamefont {R.~B.}\ \bibnamefont
  {Schnabel}},\ }\href@noop {} {\emph {\bibinfo {title} {Numerical Methods for
  Unconstrained Optimization and Nonlinear Equations}}}\ (\bibinfo  {publisher}
  {SIAM},\ \bibinfo {address} {Philadelphia},\ \bibinfo {year}
  {1996})\BibitemShut {NoStop}%
\bibitem [{\citenamefont {Vosko}\ \emph {et~al.}(1980)\citenamefont {Vosko},
  \citenamefont {Wilk},\ and\ \citenamefont {Nusair}}]{VWN}%
  \BibitemOpen
  \bibfield  {author} {\bibinfo {author} {\bibfnamefont {S.~H.}\ \bibnamefont
  {Vosko}}, \bibinfo {author} {\bibfnamefont {L.}~\bibnamefont {Wilk}}, \ and\
  \bibinfo {author} {\bibfnamefont {M.}~\bibnamefont {Nusair}},\ }\href@noop {}
  {\bibfield  {journal} {\bibinfo  {journal} {Can. J. Phys.}\ }\textbf
  {\bibinfo {volume} {58}},\ \bibinfo {pages} {1200} (\bibinfo {year}
  {1980})}\BibitemShut {NoStop}%
\bibitem [{\citenamefont {Stefanou}\ \emph {et~al.}(1987)\citenamefont
  {Stefanou}, \citenamefont {Braspenning}, \citenamefont {Zeller},\ and\
  \citenamefont {Dederichs}}]{Stefanou1987}%
  \BibitemOpen
  \bibfield  {author} {\bibinfo {author} {\bibfnamefont {N.}~\bibnamefont
  {Stefanou}}, \bibinfo {author} {\bibfnamefont {P.~J.}\ \bibnamefont
  {Braspenning}}, \bibinfo {author} {\bibfnamefont {R.}~\bibnamefont {Zeller}},
  \ and\ \bibinfo {author} {\bibfnamefont {P.~H.}\ \bibnamefont {Dederichs}},\
  }\href {\doibase 10.1103/PhysRevB.36.6372} {\bibfield  {journal} {\bibinfo
  {journal} {Phys. Rev. B}\ }\textbf {\bibinfo {volume} {36}},\ \bibinfo
  {pages} {6372} (\bibinfo {year} {1987})}\BibitemShut {NoStop}%
\bibitem [{\citenamefont {Mankovsky}\ \emph {et~al.}(2013)\citenamefont
  {Mankovsky}, \citenamefont {K\"odderitzsch}, \citenamefont {Woltersdorf},\
  and\ \citenamefont {Ebert}}]{Mankovsky2013}%
  \BibitemOpen
  \bibfield  {author} {\bibinfo {author} {\bibfnamefont {S.}~\bibnamefont
  {Mankovsky}}, \bibinfo {author} {\bibfnamefont {D.}~\bibnamefont
  {K\"odderitzsch}}, \bibinfo {author} {\bibfnamefont {G.}~\bibnamefont
  {Woltersdorf}}, \ and\ \bibinfo {author} {\bibfnamefont {H.}~\bibnamefont
  {Ebert}},\ }\href {\doibase 10.1103/PhysRevB.87.014430} {\bibfield  {journal}
  {\bibinfo  {journal} {Phys. Rev. B}\ }\textbf {\bibinfo {volume} {87}},\
  \bibinfo {pages} {014430} (\bibinfo {year} {2013})}\BibitemShut {NoStop}%
\bibitem [{\citenamefont {Anderson}(1963)}]{Anderson1963}%
  \BibitemOpen
  \bibfield  {author} {\bibinfo {author} {\bibfnamefont {P.~W.}\ \bibnamefont
  {Anderson}}\ }(\bibinfo  {publisher} {Academic Press},\ \bibinfo {year}
  {1963})\ pp.\ \bibinfo {pages} {99--214}\BibitemShut {NoStop}%
\bibitem [{\citenamefont {Rancourt}\ \emph {et~al.}(1993)\citenamefont
  {Rancourt}, \citenamefont {Dub{\'e}},\ and\ \citenamefont
  {Heron}}]{Rancourt1993}%
  \BibitemOpen
  \bibfield  {author} {\bibinfo {author} {\bibfnamefont {D.~G.}\ \bibnamefont
  {Rancourt}}, \bibinfo {author} {\bibfnamefont {M.}~\bibnamefont {Dub{\'e}}},
  \ and\ \bibinfo {author} {\bibfnamefont {P.~R.~L.}\ \bibnamefont {Heron}},\
  }\href {\doibase https://doi.org/10.1016/0304-8853(93)90816-K} {\bibfield
  {journal} {\bibinfo  {journal} {J. Magn. Magn. Mater.}\ }\textbf {\bibinfo
  {volume} {125}},\ \bibinfo {pages} {39} (\bibinfo {year} {1993})}\BibitemShut
  {NoStop}%
\bibitem [{\citenamefont {Dang}\ and\ \citenamefont
  {Rancourt}(1996)}]{Dang1996}%
  \BibitemOpen
  \bibfield  {author} {\bibinfo {author} {\bibfnamefont {M.-Z.}\ \bibnamefont
  {Dang}}\ and\ \bibinfo {author} {\bibfnamefont {D.~G.}\ \bibnamefont
  {Rancourt}},\ }\href {\doibase 10.1103/PhysRevB.53.2291} {\bibfield
  {journal} {\bibinfo  {journal} {Phys. Rev. B}\ }\textbf {\bibinfo {volume}
  {53}},\ \bibinfo {pages} {2291} (\bibinfo {year} {1996})}\BibitemShut
  {NoStop}%
\bibitem [{\citenamefont {\ifmmode \mbox{\c{S}}\else \c{S}\fi{}a\ifmmode
  \mbox{\c{s}}\else \c{s}\fi{}\ifmmode \imath \else \i
  \fi{}o\ifmmode~\breve{g}\else \u{g}\fi{}lu}\ \emph
  {et~al.}(2004)\citenamefont {\ifmmode \mbox{\c{S}}\else \c{S}\fi{}a\ifmmode
  \mbox{\c{s}}\else \c{s}\fi{}\ifmmode \imath \else \i
  \fi{}o\ifmmode~\breve{g}\else \u{g}\fi{}lu}, \citenamefont {Sandratskii},\
  and\ \citenamefont {Bruno}}]{Sasioglu2004}%
  \BibitemOpen
  \bibfield  {author} {\bibinfo {author} {\bibfnamefont {E.}~\bibnamefont
  {\ifmmode \mbox{\c{S}}\else \c{S}\fi{}a\ifmmode \mbox{\c{s}}\else
  \c{s}\fi{}\ifmmode \imath \else \i \fi{}o\ifmmode~\breve{g}\else
  \u{g}\fi{}lu}}, \bibinfo {author} {\bibfnamefont {L.~M.}\ \bibnamefont
  {Sandratskii}}, \ and\ \bibinfo {author} {\bibfnamefont {P.}~\bibnamefont
  {Bruno}},\ }\href {\doibase 10.1103/PhysRevB.70.024427} {\bibfield  {journal}
  {\bibinfo  {journal} {Phys. Rev. B}\ }\textbf {\bibinfo {volume} {70}},\
  \bibinfo {pages} {024427} (\bibinfo {year} {2004})}\BibitemShut {NoStop}%
\bibitem [{\citenamefont {Takahashi}(2013)}]{Takahashi2013}%
  \BibitemOpen
  \bibfield  {author} {\bibinfo {author} {\bibfnamefont {Y.}~\bibnamefont
  {Takahashi}},\ }\href@noop {} {\emph {\bibinfo {title} {Spin fluctuation
  theory of itinerant electron magnetism}}},\ Vol.~\bibinfo {volume} {9}\
  (\bibinfo  {publisher} {Springer},\ \bibinfo {address} {Berlin},\ \bibinfo
  {year} {2013})\BibitemShut {NoStop}%
\bibitem [{\citenamefont {Wipf}(2013)}]{Wipf2013}%
  \BibitemOpen
  \bibfield  {author} {\bibinfo {author} {\bibfnamefont {A.}~\bibnamefont
  {Wipf}},\ }\enquote {\bibinfo {title} {Mean field approximation},}\ in\ \href
  {\doibase 10.1007/978-3-642-33105-3_7} {\emph {\bibinfo {booktitle}
  {Statistical Approach to Quantum Field Theory: An Introduction}}}\ (\bibinfo
  {publisher} {Springer},\ \bibinfo {address} {Berlin},\ \bibinfo {year}
  {2013})\ pp.\ \bibinfo {pages} {119--148}\BibitemShut {NoStop}%
\bibitem [{\citenamefont {Evans}()}]{Vampire}%
  \BibitemOpen
  \bibfield  {author} {\bibinfo {author} {\bibfnamefont {R.~F.~L.}\
  \bibnamefont {Evans}},\ }\href {https://vampire.york.ac.uk} {\enquote
  {\bibinfo {title} {{VAMPIRE software package version 5.0}},}\ }\bibinfo
  {note} {Available online at \url{https://vampire.york.ac.uk}}\BibitemShut
  {NoStop}%
\bibitem [{\citenamefont {Evans}\ \emph {et~al.}(2015)\citenamefont {Evans},
  \citenamefont {Atxitia},\ and\ \citenamefont {Chantrell}}]{Evans2015}%
  \BibitemOpen
  \bibfield  {author} {\bibinfo {author} {\bibfnamefont {R.~F.~L.}\
  \bibnamefont {Evans}}, \bibinfo {author} {\bibfnamefont {U.}~\bibnamefont
  {Atxitia}}, \ and\ \bibinfo {author} {\bibfnamefont {R.~W.}\ \bibnamefont
  {Chantrell}},\ }\href@noop {} {\bibfield  {journal} {\bibinfo  {journal}
  {Phys. Rev. B}\ }\textbf {\bibinfo {volume} {91}},\ \bibinfo {pages} {144425}
  (\bibinfo {year} {2015})}\BibitemShut {NoStop}%
\bibitem [{\citenamefont {Moriya}(1985)}]{Moriya85}%
  \BibitemOpen
  \bibfield  {author} {\bibinfo {author} {\bibfnamefont {T.}~\bibnamefont
  {Moriya}},\ }\href@noop {} {\emph {\bibinfo {title} {Spin Fluctuations in
  Itinerant Electron Magnetism}}}\ (\bibinfo  {publisher} {Springer},\ \bibinfo
  {address} {Berlin},\ \bibinfo {year} {1985})\BibitemShut {NoStop}%
\bibitem [{\citenamefont {Reser}(2007)}]{Res07}%
  \BibitemOpen
  \bibfield  {author} {\bibinfo {author} {\bibfnamefont {B.~I.}\ \bibnamefont
  {Reser}},\ }\href@noop {} {\bibfield  {journal} {\bibinfo  {journal} {Phys.
  Metals Metallogr.}\ }\textbf {\bibinfo {volume} {103}},\ \bibinfo {pages}
  {357} (\bibinfo {year} {2007})}\BibitemShut {NoStop}%
\bibitem [{\citenamefont {Paradezhenko}\ \emph {et~al.}(2020)\citenamefont
  {Paradezhenko}, \citenamefont {Melnikov},\ and\ \citenamefont
  {Reser}}]{PMR20}%
  \BibitemOpen
  \bibfield  {author} {\bibinfo {author} {\bibfnamefont {G.~V.}\ \bibnamefont
  {Paradezhenko}}, \bibinfo {author} {\bibfnamefont {N.~B.}\ \bibnamefont
  {Melnikov}}, \ and\ \bibinfo {author} {\bibfnamefont {B.~I.}\ \bibnamefont
  {Reser}},\ }\href@noop {} {\bibfield  {journal} {\bibinfo  {journal} {Comp.
  Math. Math. Phys.}\ }\textbf {\bibinfo {volume} {60}},\ \bibinfo {pages}
  {404} (\bibinfo {year} {2020})}\BibitemShut {NoStop}%
\bibitem [{\citenamefont {Reser}\ \emph {et~al.}()\citenamefont {Reser},
  \citenamefont {Paradezhenko},\ and\ \citenamefont {Melnikov}}]{RPM18}%
  \BibitemOpen
  \bibfield  {author} {\bibinfo {author} {\bibfnamefont {B.~I.}\ \bibnamefont
  {Reser}}, \bibinfo {author} {\bibfnamefont {G.~V.}\ \bibnamefont
  {Paradezhenko}}, \ and\ \bibinfo {author} {\bibfnamefont {N.~B.}\
  \bibnamefont {Melnikov}},\ }\href@noop {} {\enquote {\bibinfo {title}
  {Program suite {MAGPROP 2.0}. {Federal Service for Intellectual Property
  (ROSPATENT), RU~2018617208}, 2018.}}\ }\BibitemShut {NoStop}%
\bibitem [{\citenamefont {Melnikov}\ \emph {et~al.}(2011)\citenamefont
  {Melnikov}, \citenamefont {Reser},\ and\ \citenamefont
  {Grebennikov}}]{MRG11}%
  \BibitemOpen
  \bibfield  {author} {\bibinfo {author} {\bibfnamefont {N.~B.}\ \bibnamefont
  {Melnikov}}, \bibinfo {author} {\bibfnamefont {B.~I.}\ \bibnamefont {Reser}},
  \ and\ \bibinfo {author} {\bibfnamefont {V.~I.}\ \bibnamefont
  {Grebennikov}},\ }\href@noop {} {\bibfield  {journal} {\bibinfo  {journal}
  {J. Phys. Condens. Matter}\ }\textbf {\bibinfo {volume} {23}},\ \bibinfo
  {pages} {276003} (\bibinfo {year} {2011})}\BibitemShut {NoStop}%
\bibitem [{\citenamefont {Reser}\ \emph {et~al.}(2009)\citenamefont {Reser},
  \citenamefont {Grebennikov},\ and\ \citenamefont {Melnikov}}]{Reser2009}%
  \BibitemOpen
  \bibfield  {author} {\bibinfo {author} {\bibfnamefont {B.~I.}\ \bibnamefont
  {Reser}}, \bibinfo {author} {\bibfnamefont {V.~I.}\ \bibnamefont
  {Grebennikov}}, \ and\ \bibinfo {author} {\bibfnamefont {N.~B.}\ \bibnamefont
  {Melnikov}},\ }in\ \href@noop {} {\emph {\bibinfo {booktitle} {Solid State
  Phenom.}}},\ Vol.\ \bibinfo {volume} {152}\ (\bibinfo {organization} {Trans
  Tech Publ},\ \bibinfo {year} {2009})\ pp.\ \bibinfo {pages}
  {579--582}\BibitemShut {NoStop}%
\bibitem [{\citenamefont {Fu}\ \emph {et~al.}(2019)\citenamefont {Fu},
  \citenamefont {Zhang}, \citenamefont {Duan}, \citenamefont {Dai},
  \citenamefont {Li}, \citenamefont {Xia}, \citenamefont {Jiang},\ and\
  \citenamefont {Li}}]{Fu2019}%
  \BibitemOpen
  \bibfield  {author} {\bibinfo {author} {\bibfnamefont {C.}~\bibnamefont
  {Fu}}, \bibinfo {author} {\bibfnamefont {X.}~\bibnamefont {Zhang}}, \bibinfo
  {author} {\bibfnamefont {Y.}~\bibnamefont {Duan}}, \bibinfo {author}
  {\bibfnamefont {X.}~\bibnamefont {Dai}}, \bibinfo {author} {\bibfnamefont
  {T.}~\bibnamefont {Li}}, \bibinfo {author} {\bibfnamefont {Y.}~\bibnamefont
  {Xia}}, \bibinfo {author} {\bibfnamefont {Y.}~\bibnamefont {Jiang}}, \ and\
  \bibinfo {author} {\bibfnamefont {H.}~\bibnamefont {Li}},\ }\href {\doibase
  https://doi.org/10.1016/j.jmmm.2019.165657} {\bibfield  {journal} {\bibinfo
  {journal} {J. Magn. Magn. Mater.}\ }\textbf {\bibinfo {volume} {492}},\
  \bibinfo {pages} {165657} (\bibinfo {year} {2019})}\BibitemShut {NoStop}%
\bibitem [{\citenamefont {Weinberger}\ \emph {et~al.}(2001)\citenamefont
  {Weinberger}, \citenamefont {Szunyogh}, \citenamefont {Blaas}, \citenamefont
  {Sommers},\ and\ \citenamefont {Entel}}]{Weinberger2001}%
  \BibitemOpen
  \bibfield  {author} {\bibinfo {author} {\bibfnamefont {P.}~\bibnamefont
  {Weinberger}}, \bibinfo {author} {\bibfnamefont {L.}~\bibnamefont
  {Szunyogh}}, \bibinfo {author} {\bibfnamefont {C.}~\bibnamefont {Blaas}},
  \bibinfo {author} {\bibfnamefont {C.}~\bibnamefont {Sommers}}, \ and\
  \bibinfo {author} {\bibfnamefont {P.}~\bibnamefont {Entel}},\ }\href
  {\doibase 10.1103/PhysRevB.63.094417} {\bibfield  {journal} {\bibinfo
  {journal} {Phys. Rev. B}\ }\textbf {\bibinfo {volume} {63}},\ \bibinfo
  {pages} {094417} (\bibinfo {year} {2001})}\BibitemShut {NoStop}%
\bibitem [{\citenamefont {James}\ \emph {et~al.}(1999)\citenamefont {James},
  \citenamefont {Eriksson}, \citenamefont {Johansson},\ and\ \citenamefont
  {Abrikosov}}]{James1999}%
  \BibitemOpen
  \bibfield  {author} {\bibinfo {author} {\bibfnamefont {P.}~\bibnamefont
  {James}}, \bibinfo {author} {\bibfnamefont {O.}~\bibnamefont {Eriksson}},
  \bibinfo {author} {\bibfnamefont {B.}~\bibnamefont {Johansson}}, \ and\
  \bibinfo {author} {\bibfnamefont {I.~A.}\ \bibnamefont {Abrikosov}},\ }\href
  {\doibase 10.1103/PhysRevB.59.419} {\bibfield  {journal} {\bibinfo  {journal}
  {Phys. Rev. B}\ }\textbf {\bibinfo {volume} {59}},\ \bibinfo {pages} {419}
  (\bibinfo {year} {1999})}\BibitemShut {NoStop}%
\bibitem [{\citenamefont {Mijnarends}\ \emph {et~al.}(2002)\citenamefont
  {Mijnarends}, \citenamefont {Sahrakorpi}, \citenamefont {Lindroos},\ and\
  \citenamefont {Bansil}}]{MSL02}%
  \BibitemOpen
  \bibfield  {author} {\bibinfo {author} {\bibfnamefont {P.~E.}\ \bibnamefont
  {Mijnarends}}, \bibinfo {author} {\bibfnamefont {S.}~\bibnamefont
  {Sahrakorpi}}, \bibinfo {author} {\bibfnamefont {M.}~\bibnamefont
  {Lindroos}}, \ and\ \bibinfo {author} {\bibfnamefont {A.}~\bibnamefont
  {Bansil}},\ }\href {\doibase 10.1103/PhysRevB.65.075106} {\bibfield
  {journal} {\bibinfo  {journal} {Phys. Rev. B}\ }\textbf {\bibinfo {volume}
  {65}},\ \bibinfo {pages} {075106} (\bibinfo {year} {2002})}\BibitemShut
  {NoStop}%
\bibitem [{\citenamefont {Crangle}\ and\ \citenamefont
  {Hallam}(1963)}]{Crangle1963}%
  \BibitemOpen
  \bibfield  {author} {\bibinfo {author} {\bibfnamefont {J.}~\bibnamefont
  {Crangle}}\ and\ \bibinfo {author} {\bibfnamefont {G.~C.}\ \bibnamefont
  {Hallam}},\ }\href@noop {} {\bibfield  {journal} {\bibinfo  {journal} {Proc.
  R. Soc. Lond. A}\ }\textbf {\bibinfo {volume} {272}},\ \bibinfo {pages} {119}
  (\bibinfo {year} {1963})}\BibitemShut {NoStop}%
\bibitem [{\citenamefont {Kudrnovsk\'y}\ \emph {et~al.}(2008)\citenamefont
  {Kudrnovsk\'y}, \citenamefont {Drchal},\ and\ \citenamefont
  {Bruno}}]{Kudrnovsky2008}%
  \BibitemOpen
  \bibfield  {author} {\bibinfo {author} {\bibfnamefont {J.}~\bibnamefont
  {Kudrnovsk\'y}}, \bibinfo {author} {\bibfnamefont {V.}~\bibnamefont
  {Drchal}}, \ and\ \bibinfo {author} {\bibfnamefont {P.}~\bibnamefont
  {Bruno}},\ }\href {\doibase 10.1103/PhysRevB.77.224422} {\bibfield  {journal}
  {\bibinfo  {journal} {Phys. Rev. B}\ }\textbf {\bibinfo {volume} {77}},\
  \bibinfo {pages} {224422} (\bibinfo {year} {2008})}\BibitemShut {NoStop}%
\bibitem [{\citenamefont {Evans}\ \emph {et~al.}(2014)\citenamefont {Evans},
  \citenamefont {Fan}, \citenamefont {Chureemart}, \citenamefont {Ostler},
  \citenamefont {Ellis},\ and\ \citenamefont {Chantrell}}]{Evans2014}%
  \BibitemOpen
  \bibfield  {author} {\bibinfo {author} {\bibfnamefont {R.~F.~L.}\
  \bibnamefont {Evans}}, \bibinfo {author} {\bibfnamefont {W.~J.}\ \bibnamefont
  {Fan}}, \bibinfo {author} {\bibfnamefont {P.}~\bibnamefont {Chureemart}},
  \bibinfo {author} {\bibfnamefont {T.~A.}\ \bibnamefont {Ostler}}, \bibinfo
  {author} {\bibfnamefont {M.~O.~A.}\ \bibnamefont {Ellis}}, \ and\ \bibinfo
  {author} {\bibfnamefont {R.~W.}\ \bibnamefont {Chantrell}},\ }\href@noop {}
  {\bibfield  {journal} {\bibinfo  {journal} {J. Phys. Condens. Matter}\
  }\textbf {\bibinfo {volume} {26}},\ \bibinfo {pages} {103202} (\bibinfo
  {year} {2014})}\BibitemShut {NoStop}%
\bibitem [{\citenamefont {Bruno}(2003)}]{Bruno03}%
  \BibitemOpen
  \bibfield  {author} {\bibinfo {author} {\bibfnamefont {P.}~\bibnamefont
  {Bruno}},\ }\href {\doibase 10.1103/PhysRevLett.90.087205} {\bibfield
  {journal} {\bibinfo  {journal} {Phys. Rev. Lett.}\ }\textbf {\bibinfo
  {volume} {90}},\ \bibinfo {pages} {087205} (\bibinfo {year}
  {2003})}\BibitemShut {NoStop}%
\bibitem [{\citenamefont {Zacharov}\ \emph {et~al.}(2019)\citenamefont
  {Zacharov}, \citenamefont {Arslanov}, \citenamefont {Gunin}, \citenamefont
  {Stefonishin}, \citenamefont {Bykov}, \citenamefont {Pavlov}, \citenamefont
  {Panarin}, \citenamefont {Maliutin}, \citenamefont {Rykovanov},\ and\
  \citenamefont {Fedorov}}]{Zhores}%
  \BibitemOpen
  \bibfield  {author} {\bibinfo {author} {\bibfnamefont {I.}~\bibnamefont
  {Zacharov}}, \bibinfo {author} {\bibfnamefont {R.}~\bibnamefont {Arslanov}},
  \bibinfo {author} {\bibfnamefont {M.}~\bibnamefont {Gunin}}, \bibinfo
  {author} {\bibfnamefont {D.}~\bibnamefont {Stefonishin}}, \bibinfo {author}
  {\bibfnamefont {A.}~\bibnamefont {Bykov}}, \bibinfo {author} {\bibfnamefont
  {S.}~\bibnamefont {Pavlov}}, \bibinfo {author} {\bibfnamefont
  {O.}~\bibnamefont {Panarin}}, \bibinfo {author} {\bibfnamefont
  {A.}~\bibnamefont {Maliutin}}, \bibinfo {author} {\bibfnamefont
  {S.}~\bibnamefont {Rykovanov}}, \ and\ \bibinfo {author} {\bibfnamefont
  {M.}~\bibnamefont {Fedorov}},\ }\href@noop {} {\bibfield  {journal} {\bibinfo
   {journal} {Open Eng.}\ }\textbf {\bibinfo {volume} {9}},\ \bibinfo {pages}
  {512} (\bibinfo {year} {2019})}\BibitemShut {NoStop}%
\end{thebibliography}%

\end{document}